\documentclass[final,5p,times,twocolumn,authoryear]{elsarticle}

\usepackage[authoryear]{natbib}
\usepackage{hyperref}

\usepackage{amssymb}
\usepackage{lipsum}
\usepackage{amsmath}

\journal{Astronomy $\&$ Computing}

\begin{document}

\begin{frontmatter}



\title{State Space Modelling for detecting and characterising Gravitational Waves afterglows}


\author[first,second]{D. d'Antonio}
\affiliation[first]
{organization={School of Mathematical and Physical Sciences, University of Technology, Sydney},
            addressline={15 Broadway}, 
            city={Ultimo},
            postcode={2007}, 
            state={NSW},
            country={Australia}}

\affiliation[second]
{organization={CSIRO, Space and Astronomy},
            addressline={PO Box 76}, 
            city={Epping},
            postcode={1710}, 
            state={NSW},
            country={Australia}} 


\affiliation[third]
{organization={Leonardo.Ai Research Lab},
            addressline={1 Kiara Cl}, 
            city={North Sydney},
            postcode={2060}, 
            state={NSW},
            country={Australia}}

\affiliation[fourth]
{organization={School of Mathematics and Statistics, University of Sydney},
            addressline={Carslaw Building}, 
            city={Camperdown},
            postcode={2006}, 
            state={NSW},
            country={Australia}}

\affiliation[fifth]
{organization={ARC Training Centre in Data Analytics for Resources and Environments,, University of Sydney},
            addressline={Biomedical Building}, 
            city={South Eveleigh},
            postcode={2015}, 
            state={NSW},
            country={Australia}}

\author[first,third]{M. E. Bell}
\author[first]{J. J. Brown}
\author[fourth,fifth]{C. Grazian}

\begin{abstract}
We propose the usage of an innovative method for selecting transients and variables. These sources are detected at different wavelengths across the electromagnetic spectrum spanning from radio waves to gamma-rays. We focus on radio signals and use State Space Models, which are also referred to as Dynamic Linear Models. State Space Models (and more generally parametric auto-regressive models) have been the mainstay of economic modelling for some years, but rarely they have been used in Astrophysics. 

 The statistics currently used to identify radio variables and transients are not sophisticated enough to distinguish different types of variability. These methods simply report the overall modulation and significance of the variability, and the ordering of the data in time is insignificant. State Space Models are much more advanced and can encode not only the amount and significance of the variability but also properties, such as slope, rise or decline for a given time $t$. 

In this work, we evaluate the effectiveness of State Space Models for transient and variable detection including classification in time-series astronomy. We also propose a method for detecting a transient source hosted in a variable active galaxy, whereby the time-series of a static host galaxy and the dynamic nature of the transient in the galaxy are intertwined. Furthermore, we examine the hypothetical scenario where the target transient we want to detect is the gravitational wave source GW170817 (or similar). 
\end{abstract}




\end{frontmatter}




\section{Introduction}
\label{introduction}

Transients and variables are a term for astronomical sources in the field of Time Domain Astronomy which evolve and change over time. They are largely studied across the electromagnetic spectrum and are produced from a variety of different emission mechanisms. For example, observation and discovery in the radio band has provided a great contribution to the field of time domain astronomy delivering high impact discoveries such as Fast Radio Bursts (FRBs, \citealt{Lor2007}), Pulsars \citep{Hew1968}, Gravitational Wave afterglows \citep{Abb2016} etc. 

Transients and variables have also often been studied with simultaneous radio and X-ray observing campaigns. For example, the accreting neutron stars 4U 1728-34 and 4U 1636-536 have showed correlated radio and X-ray flares \citep{Russell_2024}. 
Further discoveries in the X-ray band are particularly promising thanks to new instruments and surveys such as the Spectrum-Roentgen-Gamma (SRG)/eROSITA (\citealt{Pre2021}; \citealt{Sun2021}) All-Sky Survey which is able to provide a new eye into the X-ray transient and variable sky (\citealt{Rau2019}; \citealt{Mer2024}). 


Transients and variable sources are also studied in the gamma-ray regime. For example, Gamma-Ray Bursts (GRBs) are an example of a transient source characterised by a prompt and not repetitive gamma-ray emission \citep{Ber2014}. Optical and infrared investigations are also crucial for studying transients and variables. For instance, recent James Webb Space Telescope (JWST, \citealt{Gar2006}) observations of the gamma-ray burst GRB 221009A allowed the discovery of r-process emission \citep{Bla2023}. 

In addition, new radio telescopes and surveys can provide an accurate investigation of transient and variable radio sources due to their wide field of view. In particular, the Australian Square Kilometre Array Pathfinder (ASKAP, \citealt{ASKAP}) and the Meer Karoo Array
Telescope (MeerKAT, \citealt{Meerkat}) have already allowed the discovery of new transients (\citealt{Dri2020}, \citealt{Wang2021}). 

Machine learning algorithms have also been deployed to aid in the study of variables and transient objects. This covers a wide range of applications, for example, time series Gaussian process regression (e.g., \citealt{Boo2019}), and convolutional neural networks for image-differencing (\citealt{Her2023}; \citealt{Du2024}). Machine learning for time series processing can include Convolutional Neural Networks (CNN) and Support Vector Machines (SVM) for removing outliers from time series composed of photometric observations (e.g., \citealt{Li2024}). Also time series processing is used to aid in classifying sources of unknown classes (supernovae, gamma-ray bursts, etc.) and can integrate real-time anomaly detection with neural network classifier machines (e.g., \citealt{Lo2014}; \citealt{Gup2024}). 


In this paper, we present the use of State Space Models for better characterising the structure of variable and transient radio source time-series, with a particular focus on the gravitational wave afterglow GW170817. In subsection~\ref{intro1} we briefly describe the typical method for selecting radio variables and transients. We also define and discuss State Space Models. In subsection~\ref{why} we explain why we propose State Space Models in Astrophysics. In Section~\ref{data} we describe the data used in this work. In Section~\ref{intro_spm} different State Space Models are explained and fitted to the data. We also select the most suitable model for the data. In Section~\ref{AGN} we show a method for detecting a transient hosted by a variable galaxy and discuss variability properties. The transient we aim to detect is a gravitational wave afterglow source. Finally, in Section~\ref{conclusions} we summarise and discuss our findings.

\subsection{Variability identification and State Space Models}\label{intro1}

Time variable and transient radio sources are typically identified using metrics such as $V$ (also called modulation index $m$) and $\eta$ (also called weighted reduced $\chi^{2}$) as reported by \cite{Swi2015}, \cite{Row2019}, and \cite{Mur2021}. These metrics are defined below: 

\begin{equation}  \label{eq.1_}
\centering
   V = \frac{1}{\overline{S}} \sqrt{\frac{N}{N-1} ( \overline{S^{2}} - \overline{S}^{2}   )  },
\end{equation}

\begin{equation}  \label{eq.2_4}
 \centering \eta = \frac{N}{N-1} \left ( \overline{w S^{2}} - \frac{\overline{w S}^2}{\overline{w}}  \right ),
\end{equation}

\noindent where $N$ is the number of measurements, $S$ is the flux density value of a single measurement, $\overline{S}$ is the average flux density, $\overline{S^{2}}$ is the average square flux density, $\overline{S}^{2}$ is the square average flux density, and $w$ is the weight defined as 

\begin{equation}  \label{eq.w}
\centering
w = \displaystyle \sum_{i=0}^N \frac{1}{\sigma_{i}^{2}}, 
\end{equation}

\noindent where $\sigma_{i}$ is the error of the $i$th flux density measurement. Therefore, $\overline{wS^{2}}$ is the average of the weights multiplied by the square of flux density measurements for each source and $\overline{wS}$ is the average of the weights multiplied by the flux density measurements.  

 In a sample of sources, variables are objects having $V$ and $\eta$ (see eq.~\ref{eq.1_} and \ref{eq.2_4}) above a threshold value. For instance, \cite{Mur2021} defined variables, sources beyond a 2$\sigma$ threshold from a distribution fitted with a Gaussian function. Note that we define variables all sources which are detectable the whole observing time while transients are undetected objects and become detectable thanks to an increase in their brightness. Transients remain detectable for a limited time until they fade below the detection threshold. 

 The current strategy explained above for selecting variables has been successfully used in a number of studies (e.g., \citealt{Bell2015}, \citealt{Swi2015}, \citealt{Row2019}, \citealt{Mur2021}). However,$V$ and $\eta$ do not consider the temporal order of the data points. This is crucial as the data contain time-domain information. Time series characteristics such as slope, rise, decline and light curve trend are not characterised using the methods above. 
 
 State Space Models can identify more features in time series such as the level component in the Local Level Model (see Section~\ref{llm}) and can identify transients hosted by variable objects (see Section~\ref{AGN}). 
 
 State Space Models are time series models based on two conditions (\citealt{Tus2008}, \citealt{koller2009}):
 
 \begin{itemize}
  \item the presence of a latent or hidden process $x_{t}$ named state process at the time $t$;
  
  \item the presence of observations $y_{t}$ that are independent given the states $x_{t}$.

\end{itemize}

In general, time series models can be univariate or multivariate. The former only have one variable \citep{Brook2008} whereas the latter have more than one variable \citep{William2019}; in this paper we use Univariate State Space Models. 

These models have been widely used in different research fields such as Econometrics \citep{HAM1994}, Finance \citep{Tri2021}, Neural Data Analysis \citep{Pan2010}, Statistics \citep{jiménez2021}, Ecology and Environment (\citealt{BUCK2004}, \citealt{New2023}). Nevertheless, the particular State Space Models we discuss in this paper have been poorly applied in Time Domain Astronomy with only one application on X-ray light curves found in literature \citep{Kon1997}. 

Parametric autoregressive models have been applied in a few works (\citealt{Lazio2001}, \citealt{Tem2009}, \citealt{Kel2014}, \citealt{Fei2018}). In particular, \cite{Lazio2001} used these models to study the radio light curves of 149 radio sources at 2.5 and 8.2 GHz. The authors used autoregressive and moving average models and found that the sources analysed presented short-term variability ($\sim$10 days) caused by radio-wave scattering in an extended medium. Note that this is the only work at radio frequencies using advanced statistical time series models that we found in the literature.

We decided to use State Space Models rather than autoregressive models because of their capacity to describe hidden processes which are crucial properties in time series. The intercept in the Local Level Model is an example of a hidden process (see Section~\ref{llm}).

\subsection{Why State Space Models?}\label{why}

State Space Models provide a natural framework to identify and estimate the components of a time series such as the underlying trend, stochasticity and cyclical components \citep{Tus2008}.

These models could also be used to detect transients and variables hosted by a variable galaxy (e.g., gravitational waves source hosted by a scintillating active galaxy).

The case of the transient mentioned by \cite{Kea2016} is an example of a difficult scenario where State Space Models could be successfully used in Astrophysics. In 2015 a fast radio burst (FRB, \citealt{Lor2007}) called FRB 150418 was detected  \citep{Kea2016}. The transient was hosted by the galaxy WISE J071634.59$-$190039.2. However, other works claimed that there was no transient activity and only a scintillating active galactic nucleus (AGN) was observed (e.g., \citealt{Giro2016}; \citealt{Joh2016}). It is important to distinguish these two different scenarios and State Space Models can help to achieve this. 

State Space Models could extract the typical features of time series and resolve this kind of issue. For instance, using a state space model to fit a stochastic process would indicate a scintillating AGN (\citealt{Bhat2020}, \citealt{Sark2020}). A stochastic process is a collection of random variables ordered in time \citep{Gab2017}. A transient source model would have instead a rising phase, a peak and declining ending trend (e.g., \citealt{Dob2018}) which could in turn be fitted and extracted using a State Space Model. 

State Space Models could be used to classify transients and variables. Time domain astronomy includes the study of quite different objects such as gamma-ray bursts, flare stars, variable AGN, where each object has its own physical process driving a specific kind of time series. For example, variable AGN are likely to show stochastic light curves (e.g., \citealt{Bell2011}). Alternatively, flare stars can show light curves with bursts on the timescale of days or hours (\citealt{Ost2005}, \citealt{Fen2014}) and Gravitational Waves afterglows can show a unique bursting light curve (e.g., \citealt{Dob2018}) with individual raise and decay times. State Space Models can represent all these different behaviours, therefore these models may be used to identify and classify known and unknown transients. However, Machine Learning models are also used for classifying and detecting sources in Astrophysics (see Section~\ref{ML_SSM}). 

\subsubsection{Machine Learning vs. State Space Models}\label{ML_SSM}

Machine Learning algorithms are used in Astrophysics in several different ways such as source classification (\cite{Kar2013}; \citealt{Lo2014}; \citealt{Boo2019}; \citealt{Gup2024}), removing outliers from time series (e.g., \cite{Pow2023}; \citealt{Li2024}) and image-differencing (\citealt{Bra2008}; \citealt{Zac2016}; \citealt{San2019};   \citealt{Her2023}; \citealt{Du2024}).

Machine Learning models are  deployed for source and photometric classification. For instance, \cite{Boo2019} used Gaussian process regression for generating light curves at optical wavelengths. After training on datasets of real data light curves, the algorithm is able to generate time series over a wide range of observing conditions and redshift. The algorithm also works on poorly sampled light
curves or ones with large gaps of observations in time. Photometric classification can also be performed by adopting neural networks. For instance, \cite{Kar2013} used neural networks for classifying supernovae (SNe). The authors proposed a two-stage approach, where time series are initially fitted by an analytic parametrised function, and then, the resulting parameters are analysed by neural networks models.

Machine Learning is also used for removing light curve outliers. \cite{Li2024} used CNN and SVM for stellar and cloudy contamination identification. \cite{Pow2023} used adversarial networks (GANs) to get rid of outliers in gravitational waves sources light curves. Synthetic images were produced for 22 types of glitches commonly observed in real data and used to train the model. The neural network classification algorithm detected glitches from real data with an accuracy of 99.0\%.

Difference image analysis (DIA) is used for detecting sources and obtaining time-series photometric measurements from digital images \citep{Bra2008}. The technique allows to match each image by using a convolution kernel which takes into account changes in the point-spread function (PSF) between images \citep{Bra2008}. In addition, transient detection can also be obtained through image subtraction techniques (e.g., \citealt{Zac2016}).

Machine Learning algorithms are hence used in different ways such as sources classification (e.g., \citealt{Kar2013}) and transient detection through DIA (e.g., \citealt{Zac2016}). However, in Section~\ref{why}, we mention that State Space Models may also be utilised for classifying variables and transients. The advantage of State Space Models is the nonessential need of training large datasets. 

State Space Models may also provide other use cases as we mention in Section~\ref{why}. In particular, State Space Models can extract variability components from time series (e.g., stochasticity, light curve trend) and detect transients hosted by a variable galaxy. The extraction of variability components and the transient detection method are to be described in Section~\ref{intro_spm} and \ref{AGN}, respectively.

\section{Data}\label{data}

We applied State Space Models to the radio afterglow of the gravitational waves event GW170817 which was the first afterglow of a gravitational wave source detectable by electromagnetic telescopes \citep{Abb2017}.  In particular, we used the data from the follow-up observations of GW170817 carried out by the Australian
Square Kilometre Array Pathfinder (ASKAP, \citealt{ASKAP}). The observations are from \cite{Hal2017}, \cite{Moo2018a}, \cite{Dob2018}, \cite{Mooley2018c}, \cite{Alexander2018}, \cite{Margutti2018}, and \cite{Troja_2019}. 

\cite{Dob2019} explain that follow-up observations of GW170817 started 15 hours after the event by searching for coherent radio emission in fly’s-eye mode (\citealt{Ban2017b}; \citealt{Ban2017c}) while imaging observations began two days after the event \citep{Dob2019}.

We used the light curve data points reported by \cite{Dob2019} that scaled the ASKAP follow-up observations to 1.4 GHz using a spectral index $\alpha$ = -0.58. In detail, we used the power-law commonly used in radioastronomy and reported by \cite{astro_bible}: 

\begin{equation}  \label{index}
\centering
I(\nu) = I(\nu_{0}) \left (\frac{\nu} {\nu_{0}}\right)^{\alpha}, 
\end{equation}

\noindent where $I(\nu)$ and $I(\nu_{0})$ are flux densities at the frequency $\nu$ and $\nu_{0}$, respectively and $\nu < \nu_{0}$. Here $\alpha$ is the spectral index and for this relationship sources are brighter at lower frequencies. 

\section{Testing State Space Models on GW170817}\label{intro_spm}

In this section, we describe the method adopted to fit State Space Models on the GW170817 light curve. We show the different features of each model. We also select the most suitable model for fitting the data.

\subsection{Introduction on model selection criterion and heteroskedasticity}

To select the best model for describing the GW170817 light curve, we used several goodness-of-fit statistics which are often implemented in time series. These statistics are the Aikaike Information Criterion (AIC), the Bayesian Information Criterion (BIC), the Hannan-Quinn Information Criterion (HQIC) and the Heteroskedasticity (H). 

The AIC provides a measurement of the model ``goodness-of-fit'' with respect to the data and is defined by the following formula \citep{Ken2004}: 
\begin{equation}  \label{AIC}
\centering
  \textrm{AIC} = 2k - 2\ln{L},
\end{equation}
\noindent where $L$ is the  maximised likelihood function which is the Joint Probability of the observations as a function of the parameters of the statistical model. Suppose we have a sample of observations $y_{t}$. The probability density function (PDF) of each observation is $f(y_{i}| \theta)$ where $\theta$ is a parameter of the statistical model. Assuming independent observations, the likelihood function is given by 
\begin{equation}  \label{L}
\centering
  L(\theta) = f(y_{1}| \theta) \cdot f(y_{2}| \theta)...\cdot f(y_{n}| \theta) = \prod_{n=1}^{N} f(y_{i}|\theta);
\end{equation}
We notice that $\theta$ can be a scalar or a vector, depending on the adopted model. In our case the observations are the flux density values of GW170817 light curve, and we define the dimension of $\theta$ as $k$. 

The BIC is also called the Schwarz Information Criterion. It is a model selection criterion (likewise the AIC) and is defined by the relation \citep{Ken2004}:
\begin{equation} \label{BIC}
\centering
   \textrm{BIC} = k\ln{N} - 2\ln{L},
\end{equation}
where $N$ is the sample size.

Another alternative for selecting models is the HQIC. This is formally defined as \citep{Bur2002}: 
\begin{equation} \label{HQUIC}
\centering
   \textrm{HQIC} = - 2L + 2k\ln({\ln{N}}).
	\end{equation}
Lower values of AIC, BIC and HQIC implies a better statistical model to represent the observed data. In this work, we compared the values of these three parameters among different models.

The Heteroskedasticity \citep{Bar2006} gives an estimation of the conditional variance ($Variance (y|X)$) which is the variability of the observed data $y_{t}$ for each value of the variable $X$ or time $t$. We have a heteroscedastic data set when the standard deviations $\sigma$ of a predicted variable $y$ are not constant over an independent variable $X$ or time $t$. This implies that the absolute residuals of the variable $y$ are not constant over the variable $X$ or the time $t$. The residuals are the difference between the predicted values and the actual measurements of the variable $y$. If the standard deviations and the absolute residuals were constant, we would see a homoskedastic data set. 

Notice that unlike AIC, BIC, and HQIC, the heteroskedasticity is not a model selection criterion but gives a measure of the data points variance over time. In particular, heteroskedasticity is interesting as it can impact the model fitting. In simple terms, it is more difficult to fit a time series with variance changing over time rather than fitting a time series with constant variance \citep{pes2015}. 

\subsubsection{Testing time series stationarity }\label{stationary}
Before trying to fit the GW 170817 light curve with statistical models, it was necessary to find out if the time series is stationary. Some models are, in fact, suitable to stationary time series only (see Section~\ref{arima}). A time series is stationary if the following conditions are satisfied \citep{Kwi1992}:
\begin{itemize}
    \item constant mean $\mu$ over time $t$;
    \item constant variance $\sigma$ over time $t$;
    \item the Autocorrelation function (ACF) have a steep decline to 0.0. 
\end{itemize}

\begin{figure}
       \hspace*{-0.5cm} 
		\includegraphics[scale=0.35]{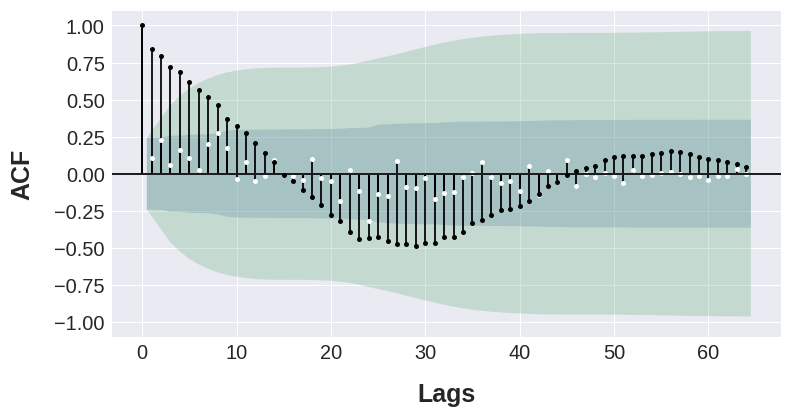}
		\caption{ \small {\label{ACF} {Autocorrelation functions of GW170817 light curve and a stationary time series. The black points are the ACF of GW170817 while the white points are the ACF of stationary time series. The green shaded area is the confidence interval of the ACF of GW170817. The blue shaded area is the confidence interval of the stationary time series ACF. There is a probability of 95\% to find a correlation within each confidence region.}}}
\end{figure}

To verify whether if GW170817 time series is stationary, we analysed its Autocorrelation function (ACF). This is a function giving the correlation between the value of the time series at a given time $t$ and the value of the same time series at time $t-1$. In other words, we see the correlation between two flux density values $y_{t}$ and $y_{t-1}$. The formula of autocorrelation for a time length of observation equal to $T$ is reported below \citep{shum2017}: 
\begin{equation} \label{autocorrelation}
\centering
   \textrm{ACF}(h) = \frac { \sum_{t=1}^{T-h} (y_{t+h} - \bar{y}) (y_{t} - \bar{y }) } {T},
\end{equation}
where $y_{t}$ is the flux density value at the time $t$, $\bar{y}$ is the mean flux density while $h$ is the lag. Every lag is the time difference between $y_{t}$ and $y_{t+h}$. For $h = 1$ the autocorrelation is estimated between $y_{t-1}$ and $y_{t}$ while for $h = 2$ it is estimated between $y_{t-2}$ and $y_{t}$ and so on. Stationary time series have ACFs with a dramatic decline around 0.0 and maintain this value for all lags \citep{shum2017}. In Fig~\ref{ACF} we can compare the ACF of GW170817
with the ACF of a stationary time series. Whilst the stationary time series ACF declines to 0.0 by increasing the number of lags, the ACF of GW170817 shows a different behaviour as does not keep a value near 0.0 for all lags (see Fig~\ref{ACF}). We thus concluded the light curve of GW170817 is not stationary. 

\subsection{Fitting GW170817 light curve with State Space Models}

\subsubsection{Local Level Model}\label{llm}
The \textbf{Local Level Model} is expressed with the two equations below \citep{Tus2008}: 

\begin{equation}  \label{yt}
\centering
y_{t} = \mu_{t} + \epsilon_{t},  \qquad \epsilon_{t} \sim N(0,\sigma_{\epsilon}^{2}),
\end{equation}

\begin{equation}  \label{mut}
\centering
\mu_{t+1} = \mu_{t} + \xi_{t},  \qquad \xi_{t} \sim N(0,\sigma_{\xi}^{2}),
\end{equation}

\noindent where $y_{t}$ is the observation at the given time $t$. If we were analysing the light curve of an astronomical source, $y_{t}$ would be the flux density. The \textit{level component} $\mu_{t}$ can be seen as the intercept of the function over time. The term $\epsilon_{t}$ is an \textit{irregular component} that is an observation disturbance. It represents the error to add on the signal. Finally, $\xi_{t}$ is called the level disturbance as it can be seen as the error associated with the \textit{level component}. The two disturbance terms have a normal distribution centered at 0 and with specific variance term, as shown by the notation $N(0,\sigma^{2})$ in equations \eqref{yt} and \eqref{mut}. 

The Local Level Model satisfies the two criteria required to have a State Space Model: the presence of a latent process and conditionally independent observations \citep{Tus2008}.  
The flux density measurements $y_{t}$ are the observations which are measured directly. The latent process is given by $\mu_{t}$ which is the intercept for each value of flux density at a given time $t$. We do not measure this term directly. 

\begin{figure*}
		\includegraphics[scale=0.53]{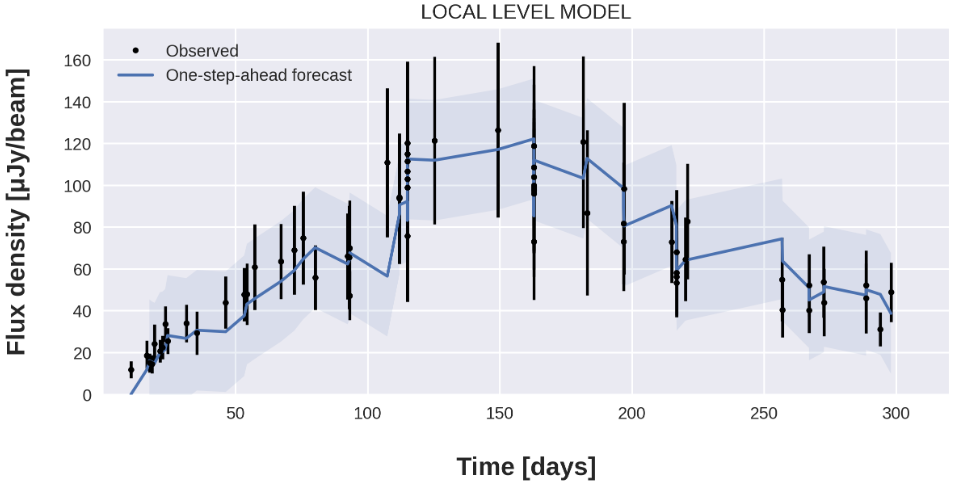}
		\caption{ \small {\label{local} {Local level model fitted to the gravitational wave event GW170817 \citep{Dob2018}. The black points (with their errors also in black) are data from \cite{Dob2018}. The blue line is the modelled fit of the light curve. The light blue area is the 95\% confidence region. }}}
\end{figure*}

We can use this model to describe a light curve as shown in Fig. \ref{local}, where the gravitational wave event GW170817 \citep{Dob2019} and the corresponding fitted morel are shown. We added a 95\% confidence region. Every modelled observation at a given time $t$ has been estimated based on the value of the previous one at the time $t-1$. Most measured data points are nearby the blue line of the model and inside the confidence region, therefore, this is a good fit. 

This model can hence be used to find other transients with the same physical origin and to extract physical parameters from the time series. For example, we could extract the gradient to derive the rise phase gradient and thus the physical parameters of the explosion / merger. However, this model may have limitations on extracting several physical parameters. The Local Level Model is essentially based on one component which is the $level$ $component$. Extracting several physical components from one single model component only may require a further development of the model itself.

\subsection{Local Linear Trend Model}\label{LLT}

By adding a slope term $\nu_{t}$ to the  Local Level Model, we obtain the Local Linear Trend Model \citep{Koop2012}: 

\begin{equation}  \label{y}
\centering
y_{t} = \mu_{t} + \epsilon_{t},  \qquad \epsilon_{t} \sim N(0,\sigma_{\epsilon}^{2}),
\end{equation}

\begin{equation}  \label{mu}
\centering
\mu_{t+1} = \mu_{t} + \nu_{t}  + \xi_{t},  \qquad \xi_{t} \sim N(0,\sigma_{\xi}^{2}),
\end{equation}

\begin{equation}  \label{mu}
\centering
\nu_{t+1} = \nu_{t} + \zeta_{t},  \qquad \zeta_{t} \sim N(0,\sigma_{\zeta}^{2}),
\end{equation}

\begin{equation}  \label{slope}
\centering
\nu_{t} = (y_{t}+w_{t})t,    \qquad w_{t} \sim N(0,\sigma_{w}^{2}),
\end{equation}

\noindent the term $\nu_{t}$ is a slope term generated by a random walk and defined by eq. \eqref{slope}. A random walk is a time series model where the next observation $y_{t+1}$ equals the previous one $y_{t}$ with a random step up or down \citep{Koop2012}. Moreover, $\epsilon_{t}$, $\xi_{t}$, $\zeta_{t}$, and $w_{t}$ are disturbance terms normally distributed. 

\begin{figure*}
        \hspace*{-0.3cm} 
		\includegraphics[scale=0.53]{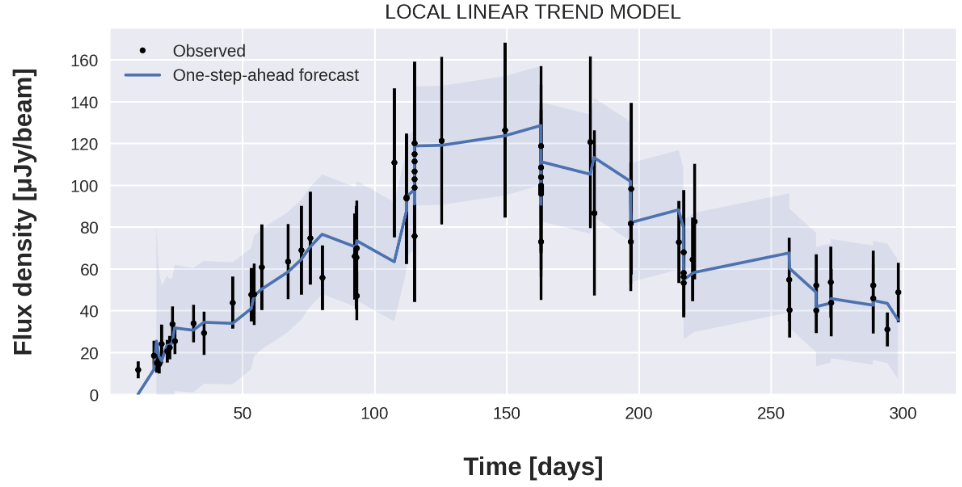}
		\caption{ \small {\label{local_trend} {Local Linear Trend Model fitted to the gravitational wave event GW170817 \citep{Dob2018}. The black points (with their errors also in black) are data from \cite{Dob2018}. The blue line is the modelled fit of the light curve. The light blue area is the 95\% confidence region. }}}
\end{figure*}

In Fig.~\ref{local_trend} we can see the model fitting the data. This fitting may appear to be very similar to the Local Level Model one, however, the average confidence region of the Local Linear Trend Model is wider (see Table~\ref{conf_reg}).

\subsection{Autoregressive State Space Model}\label{AR}

Autoregressive (AR) models can be written as \citep{Fei2018}: 

\begin{equation} \label{AR}
\centering
y_{t} = a_{1}y_{t-1} + a_{2}y_{t-2}+...+a_{p}y_{t-p} + \epsilon_{t},
\end{equation}

\noindent where $y_{t}$ indicates the observed data, $a_{1}$, $a_{2}$....$a_{p}$ are coeffients, $\epsilon_{t}$ is a normally distributed random error and $p$ is the order of the model. If $p = 2$, the model is: 

\begin{equation} \label{AR}
\centering
   y_{t} = a_{1}y_{t-1} + a_{2}y_{t-2} + \epsilon_{t}, \qquad \epsilon_{t} \sim N(0,\sigma^{2})
\end{equation}

\begin{figure*}
       \hspace*{-0.6cm} 
		\includegraphics[scale=0.53]{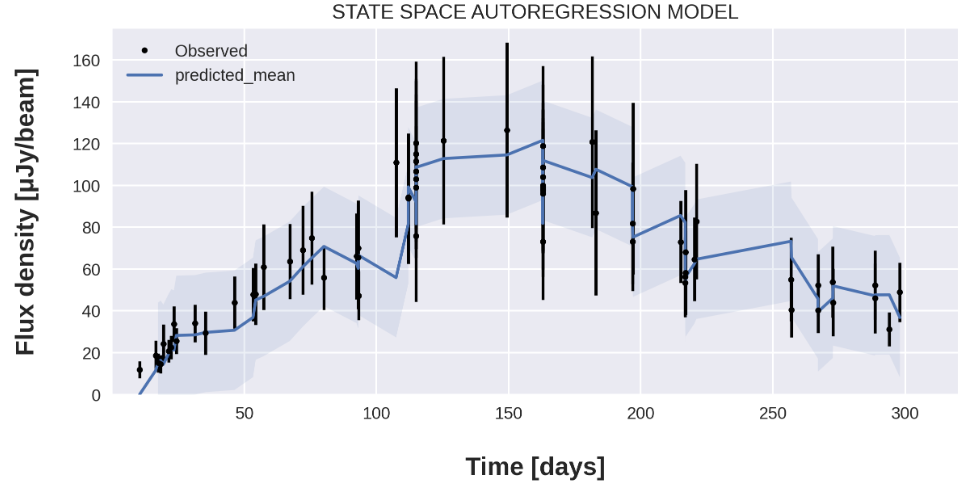}
		\caption{ \small {\label{AR2} {AR(2) model in state space form fitted to the gravitational waves event GW170817 \citep{Dob2018}. The black points (with their errors also in black) are data from \cite{Dob2018}. The blue line is the modelled fit of the light curve. The light blue area is the 95\% confidence region.   }}}
\end{figure*}

\noindent In space state form the model is defined as below:

\begin{equation}
y_{t} = \alpha_{t} 
\begin{vmatrix} 1&0 \end{vmatrix},
\end{equation}

\begin{equation}
\alpha_{t+1} = 
 \begin{vmatrix} 
 a_{1}&a_{2}\\1&0   
 \end{vmatrix}  
 \alpha_{t} + 
 \begin{vmatrix} 
 1\\0  
 \end{vmatrix} 
 \eta_{t}, \qquad \eta_{t} \equiv \epsilon_{t+1} \sim N(0,\sigma^{2}), 
\qquad 
\end{equation}

\noindent where $\eta_{t}$ is a disturbance term normally distributed. The model with $p = 2$ is able to describe the data efficiently (see Fig.~\ref{AR2}). We have decided to fit an autoregressive model with $p=2$, as this was the value for which we found the lowest AIC, BIC, and HQIC for the autoregressive model. 


\subsection{State Space ARIMA Model}\label{arima}

In the GW170817 time series there is a rising trend up to 149 days \citep{Dob2018} and then a falling behaviour up to 300 days \citep{Dob2019}. However, in the rising phase of the light curve (up to 149 days) we do not see a continuing rise fo all data points $y_{t} > y_{t-1}$. Every now and then, the light curve rises, declines, rise again and so on with many more rises than declines. The light curve is thus rising up to 149 days. When in statistical time series analysis, we see this combination of rises and declines, we talk about ``shocks''. In the GW light curve, there are obviously shocks between 150 and 300 days. 

ARMA(p,q) models contain an autoregressive (AR) process and a moving average (MA) process \citep{Fei2018}, and can be written as:
\begin{equation} \label{AR}
\centering
   y_{t} - a_{1}y_{t-1} + a_{2}y_{t-2}+...+a_{p}y_{t-p} = \epsilon_t + b_1 \epsilon_{t-1} + \ldots + b_q \epsilon_{t-q},
\end{equation}
where $y_{t}$ indicates the observed data, $a_{1}$, $a_{2}$....$a_{p}$ are AR coeffients, $\epsilon_{t}$ is a normally distributed random error, and $b_1, b_2, \ldots, b_q$ are MA coefficients. The error terms $\epsilon_{t-1}, \epsilon_{t-2}, \ldots, \epsilon_{t-q}$ are called ``random shocks''. In general, we talk about ARMA (p,q) process, where p and q are the orders of the AR and the MA process, respectively. ARMA models are suitable for time series with the assumption of stationarity. This condition does not apply to GW170817 light curve (see Section~\ref{stationary}).  

It is possible to remove the non-stationarity by differencing \citep{Fei2018}. In fact, we can replace the time series $y_{t}$ with another $y'_{t}$ that can be written as:
\begin{equation} \label{ARIMA}
\qquad \qquad
   y'_{t} = y_{t} - By_{t} = y_{t} - y_{t-1},
\end{equation}
where $B$ is called backshift (or lag) operator \citep{Fei2018}. We can use a stationary ARMA process for the differenced time series, instead of the original time series. The original time series can be recreated by reversing or integrating the differenced time series. This process is called ARIMA(p,d,q) model where d is the number of differencing operations applied to the original time series \citep{Fei2018}. In eq.~\eqref{ARIMA} this parameter is d $= 1$. 

The model we used is an ARIMA model of order d=2 in state space form \citep{Dur2012}: 
\begin{equation} \label{st_arima1}
\centering
   \Delta^{2} y_{t} = \Delta( y_{t} -  y_{t-1}),
\end{equation}

\begin{equation} \label{st_arima2}
\centering
   y_{t} = \Delta y_{t} + y_{t-1} = \Delta^{2} y_{t}+ \Delta y_{t-1} + y_{t-1},
\end{equation}
where $\Delta$ is the differencial operator. The hidden process is given by eq.~\eqref{st_arima1} and describes the flux density gradient over time. The model that we applied is a State Space ARIMA (1,2,1) which is also called SSARIMA(1,2,1). We chose the values of the three parameters (p,d,q) by comparing the AIC, BIC and HQIC for different ARIMA models with various configurations of (p,d,q). The values tested of p,d, and q  were all the integer values from 0 to 20. 

\begin{figure*}
        \hspace*{-0.3cm} 
		\includegraphics[scale=0.53]{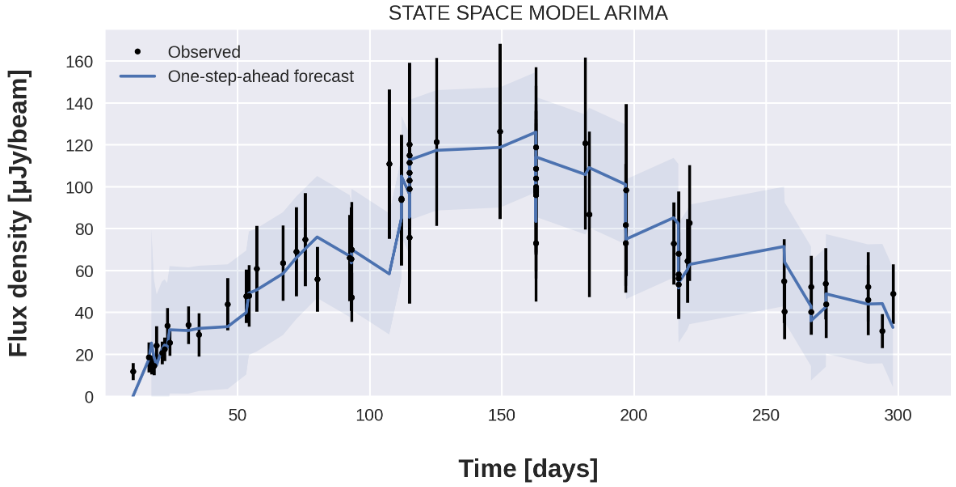}
  \caption{\small {\label{state_space_arima}ARIMA(1,2,1) model in state space form fitted to the gravitational wave event GW170817 \citep{Dob2018}. The black points (with their errors also in black) are data from \cite{Dob2018}. The blue line is the modelled fit of the light curve. The light blue area is the 95\% confidence region. }}
\end{figure*}

In Fig~\ref{state_space_arima} we can see that the model properly fits the observed data. 

\subsection{Best model selection}\label{best}

We tested four state space models: Local Level Model (LLM), Local Linear Trend (LLT), State Space Autoregression (AR) Model, and SSARIMA (1,2,1). 

In Table~\ref{conf_reg} the width average, the maximum width, the minimum width and the width median of the confidence region for each model are reported. The four models have confidence regions with similar size. Note that we estimated these mean values without considering the first data point. Because in these models every value $y_{t}$ is modelled considering the previous one $y_{t-1}$, the first data point always has a large confidence region which is not statistically relevant.

\begin{table}
\caption{\label{conf_reg} Confidence region width of the five models adopted: Local Level Model (LLM), Local Linear Trend (LLT) Model, State Space AR(2) Model, SSARIMA (1,2,1) and SSARIMA(2,1,1) with missing values. The first data point over time was not included for estimating these parameters.  }
\centering
\renewcommand{\arraystretch}{1.3}
\begin{tabular}{c | c | c | c | c }  
\hline \hline
Model &  LLM &  LLT  & AR(2)  & SSARIMA(1,2,1)     \\ 
\hline
Width average   & 56.6  & 57.5    & 55.9    & 57.7 \\ 
Max width   & 57.6  & 79.8    &  57.1   & 79.4    \\
Min width   & 43.8 &  51.9  & 43.0    &  48.6   \\
Width median & 57.6   &  57.0    & 57.1 & 57.3  \\
\hline
\end{tabular}
\end{table}

The model selection criteria and the heteroskedasticity values are reported in Table~\ref{global}. The Local Level Model is the model associated with the lowest values of all goodness-of-fit measures. This indicates that the Local Level Model is the most suitable model for the observed data. However, before establishing that the Local Level Model is the best model, it is necessary to run detailed tests on the model residuals. This is explained in Section~\ref{model_diagnostics}. 

\begin{table}
\caption{\label{global} Goodness-of-fit measures of the four models adopted: Local Level Model (LLM), Local Linear Trend (LLT) Model, State Space AR(2) Model and SSARIMA (1,2,1). The letter H stands for heteroskedasticity.}
\centering
\renewcommand{\arraystretch}{1.3}
\begin{tabular}{c | c | c | c | c }  
\hline \hline
Model &  LLM &  LLT  & AR(2)  & SSARIMA(1,2,1)      \\    
\hline 
AIC &  531.8     &  533.9   & 549.9    & 536.3  \\ 
BIC & 538.2  &  540.4   & 556.5    & 545.0   \\
HQIC & 534.3  & 536.4   & 552.5     & 539.7    \\
H  & 2.6  & 2.8   & 3.5     & 3.3 \\
\hline
\end{tabular}
\end{table}

\subsection{Model diagnostics}\label{model_diagnostics}

In this Section, we offer additional tests based on residual diagnostics of fit for the Local Level Model compared with the other available models (see Section~\ref{best}). 

Residuals in Univariate Space State Models are supposed to satisfy three properties \citep{Tus2008}: 

\begin{itemize}
    \item independence;
    \item homoskedasticity;
    \item normality.
\end{itemize}

If at least one of these properties is not verified, other models may be more suitable. 

\begin{figure}
        \hspace*{0.3cm} 
		\includegraphics[scale=0.3]{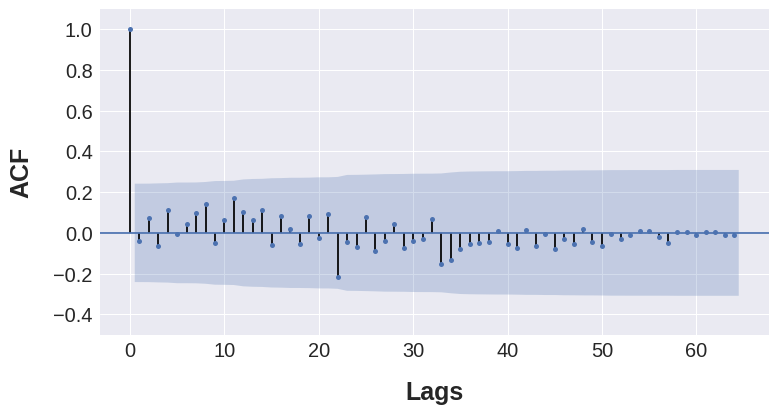}
		\caption{ \small {\label{autocorrelation_res} {Autocorrelation function of the residuals from the Local Level Model. The blue area is the confidence region which is included in the interval $\pm1.96 \sqrt{N}$. }}}
\end{figure}

We verified that the residuals are independent with the plot of the Autocorrelation function of the residuals in Fig.~\ref{autocorrelation_res}.  The values fall in the confidence region of the autocorrelation function which is given by the limits $\pm1.96 \sqrt{N}$. When the autocorrelations are within this confidence region, the mean of the residuals is enough close to zero to state that there is no evident correlation in the residuals time series \citep{Broc2010}.

\begin{figure}
        \hspace*{0.3cm} 
		\includegraphics[scale=0.3]{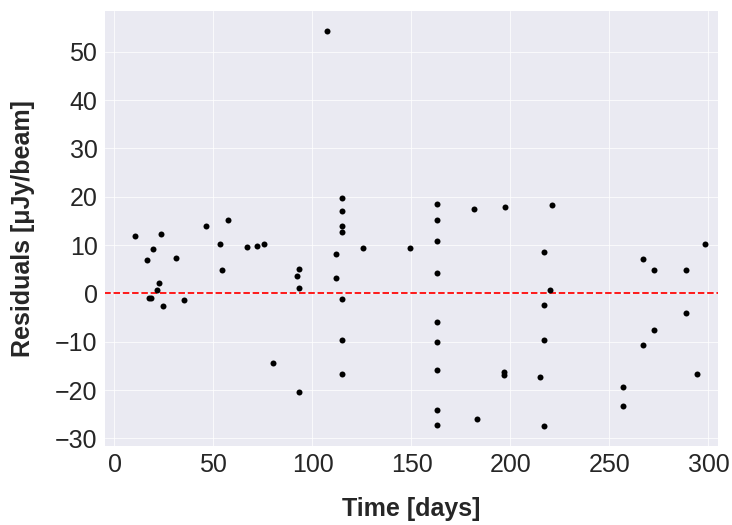}
		\caption{ \small {\label{LLM_res_time} {Residuals vs. time for the Local Level Model.  }}}
\end{figure}

To satisfy the condition of homoskedasticity it is necessary to have residuals with constant variance. A simple check is a plot showing the residuals vs. time or responses. If the condition of homoskedasticity is satisfied, the spread of residuals vs. time needs to be roughly constant. The dispersion around 0 seems to increase from a time of 80 days (see Fig.~\ref{LLM_res_time}). This is due to the presence of a residual value above 50. Apart from this large residual, all the other values are between -25 and 20. The trend is roughly constant. Residuals vs time were plotted even for the other time series models explored in this work. The other models also showed a similar behaviour with a residual value much larger than all the others. 

The data point causing this large residual was observed at 107.4 days in the GW170817 light curve. The data point is not included in the confidence region of the Local Level Model (see Fig.~\ref{local}). Interestingly, this ``critical'' data point is an outlier for all the time series models considered in this work (see also Fig.~\ref{local_trend}, \ref{AR2}, \ref{state_space_arima}). This data point is above 100 $\mu$Jy and starts a sudden raise in the time series as the previous data point is below 80 $\mu$Jy. In general, the one-step-ahead models examined in this work, struggle on predicting sudden rises or declines. The observations following the critical data point are properly fitted by the all time series models. 

\begin{figure}
        \hspace*{0.85cm} 
		\includegraphics[scale=0.35]{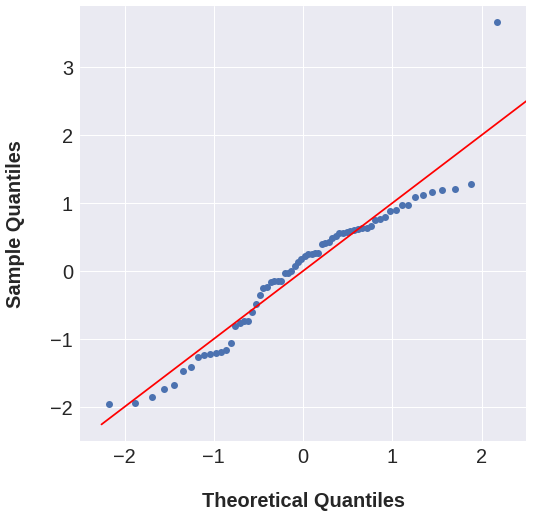}
		\caption{ \small {\label{qq} {Q-Q plot of the residuals.    }}}
\end{figure}

We analysed the Q-Q plot to verify that the residuals follow a normal distribution (see Fig~\ref{qq}). 
 A Q-Q plot is a graph of theoretical quantiles of a normal distribution with mean $\mu$ = $0$ and standard deviation $\sigma$ = 1 \citep{Koop2012}. We also plotted the sample quantiles. In this case the sample quantiles are the residuals. If the points of the Q-Q plot fall on the diagonal $y=x$, the residuals follow a normal distribution \citep{Koop2012}. Despite a small spread, most points on the Q-Q plot follow this trend except the one with the largest residual (see Fig~\ref{qq}). We hence concluded that the residuals approximately follow a normal distribution.    
 
In summary, the residuals of the Local Level Model satisfied the conditions of independence, normality and homoskedasticity. Thus, the conclusion is that the Local Level Model is suitable for the GW170817 light curve. 

\section{A method for detecting a transient hosted by an active galaxy}\label{AGN}

In this section we show how State Space Models can
detect a transient or gravitational waves event within a variable galaxy. Distinguishing between
a transient signal and an active galaxy signal could be a complex task if the transient
is within a variable active galaxy. We created a scenario where GW170817 is associated with a
simulated AGN light curve and we used State Space Models to detect the burst from the transient source.

\subsection{Testing State Models on a simulated AGN}

\begin{figure*}
		\includegraphics[scale=0.51]{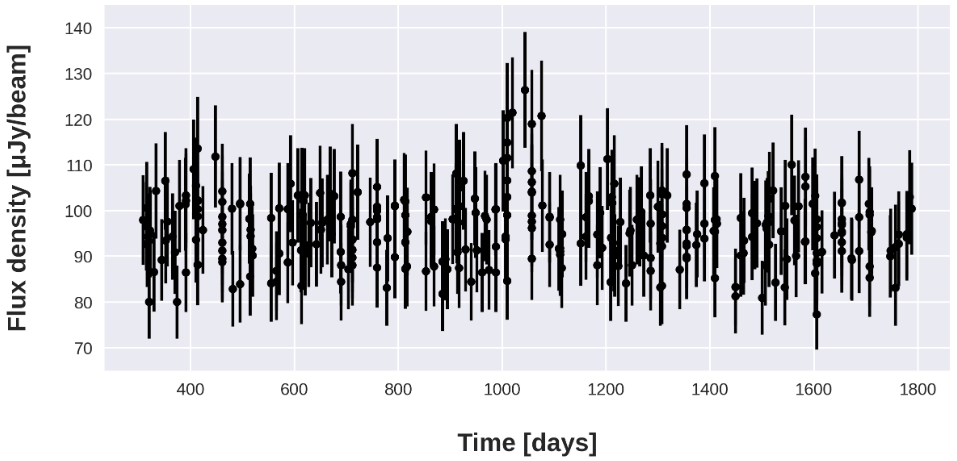}
		\caption{ \small {\label{GW_AGN2} {GW170817 hosted by a bright AGN. The transient source is between 1000 and 1200 days of observations. However, the presence of the source is not evident as it is within a variable active galaxy.  }}}
\end{figure*}

We built a ``fake'' AGN light curve of synthetic data (see Fig.~\ref{GW_AGN2}) with the same sampling of the GW (gravitational waves) time series extended by five cycles. The synthetic data are roughly between 70 and 120 $\mu$Jy with a mean value of 95 $\mu$Jy. Each associated error bar is 10\% of its flux density measurement. By adding GW170817 to the AGN light curve, we got a mean value of 96.1 $\mu$Jy which is the average flux density of the AGN in the host galaxy of GW170817. In detail, we scaled to 1.4 GHz the real data from \cite{Ban2017} at 16.7 GHz and \cite{Ale2017} at 9.77 GHz. The average of this real data is 96.1 $\mu$Jy.  We chose these flux density values to make sure the AGN was not excessively bright and the mean flux density was not extremely low. With an excessively bright AGN, a transient would not be detectable. On the other hand, with an excessively faint AGN a transient would be very easily detectable. We aimed for a scenario for which transient activity may be within an AGN light curve but the detection would be unclear. We have basically reproduced a scenario similar to the case of FRB 150418 (\citealt{Kea2016}; \citealt{Joh2016}).

We fitted different time series models on the AGN light curve with the same approach adopted in Section~\ref{intro_spm}. The models are the following: Local Level Model (LLM), Local Linear Trend (LLT) Model, State Space AR(3) Model, and SSARIMA(3,1,25) and the result of their statistical performance is reported in Table~\ref{global2}. We tested different values of p,d, and q for the State Space AR Model and SSARIMA. We tried integer values of p from 0 to 50 for Model State AR and integer values of p,d, and q from 0 to 50 for SSARIMA.  

\begin{table}
\caption{\label{global2} Statistical results of the four models tested: Local Level Model (LLM), Local Linear Trend (LLT) Model, State Space AR(3) Model and SSARIMA (3,1,25). The letter H stands for heteroskedasticity.  }
\centering
\renewcommand{\arraystretch}{1.3}
\begin{tabular}{c | c | c | c | c }  
\hline \hline
Model &  LLM &  LLT  & AR(3)  & SSARIMA(3,1,25)      \\
       
\hline 

AIC & 2259.1  &  2270.2   &  2264.2   &  2285.4 \\ 
 
BIC & 2270.5  &  2281.6  &  2283.2   &   2399.3\\

HQIC & 2263.7  &  2274.7  &   2271.8   &    2330.8\\
 
H  & 0.9   & 0.9    &  0.9    &  0.8 \\

\hline
\end{tabular}
\end{table}

\begin{figure*}
		\includegraphics[scale=0.51]{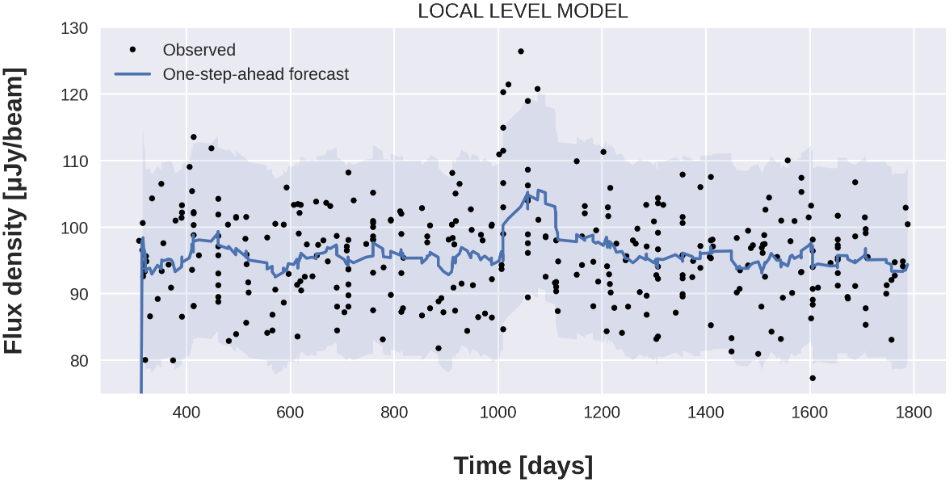}
		\caption{ \small {\label{LLM_AGN} {Local Level Model fitting the light curve. The error bars were omitted to clearly show the estimated mean curve.  }}}
\end{figure*}

The Local Level Model is the model with the best parameters (see Table~\ref{global2}). In Fig.~\ref{LLM_AGN} we can see the model fitting the time series. For simplicity, in this Section we only show the fit of the best time series model. 

\begin{figure*}
		\includegraphics[scale=0.51]{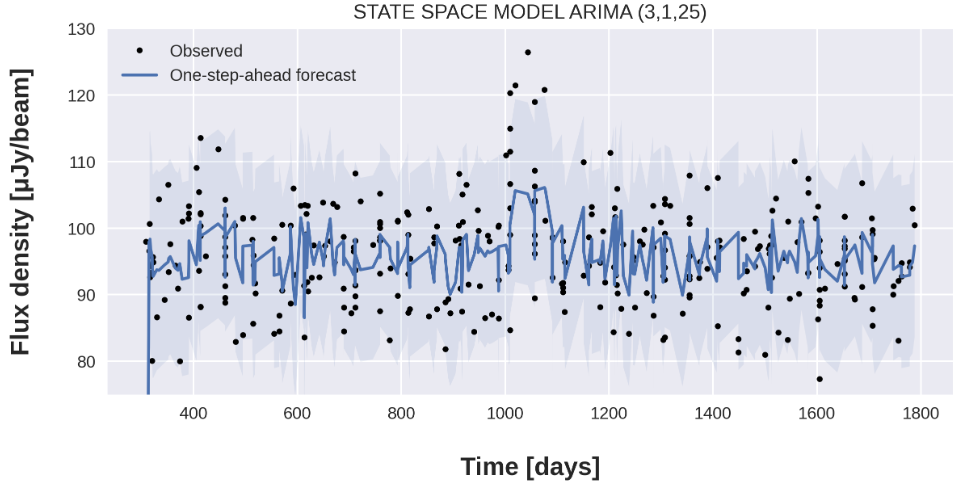}
		\caption{ \small {\label{AGN_SSARIMA} {SSARIMA(3,1,25) model fitting the light curve. The light blue region is the 95\% confidence region. The error bars were omitted to clearly show the estimated mean curve.  }}}
\end{figure*}

\subsubsection{Analysing stochastic behaviour} 

Interestingly, the SSARIMA model can reproduce the stochastic behaviour of the time series (see Fig.~\ref{AGN_SSARIMA}).  

Stochastic time series have several declines and rises between consecutive data points. ARIMA models are suitable to reproduce this behaviour. Rises and declines can indeed be fitted using a regression and a moving average process (see Section~\ref{arima}). This is demonstrated by works in literature showing ARIMA models fitting AGN light curves (\citealt{Bhat2020}, \citealt{Sark2020}). In this case, the high moving average of 25 reproduces the local shocks in the time series while the regression order of 3 reproduces the global stochastic trend of the data. Hence, a SSARIMA model can highlight the stochastic variability of time series. 

Notice that the values of the moving average and regression orders were chosen after testing integer values from 0 to 50. The values of 25 and 3 gave the lowest goodness-of-fit measures (AIC, BIC, HQIC). 

Despite the SSARIMA model describes stochastic trends, the Local Level Model is the best choice for fitting the AGN time series (see Table~\ref{global2}). This is why we used this model for the transient detection method explained in Section~\ref{detection}.

\subsection{Transient detection}\label{detection}

\begin{figure*}
		\includegraphics[scale=0.53]{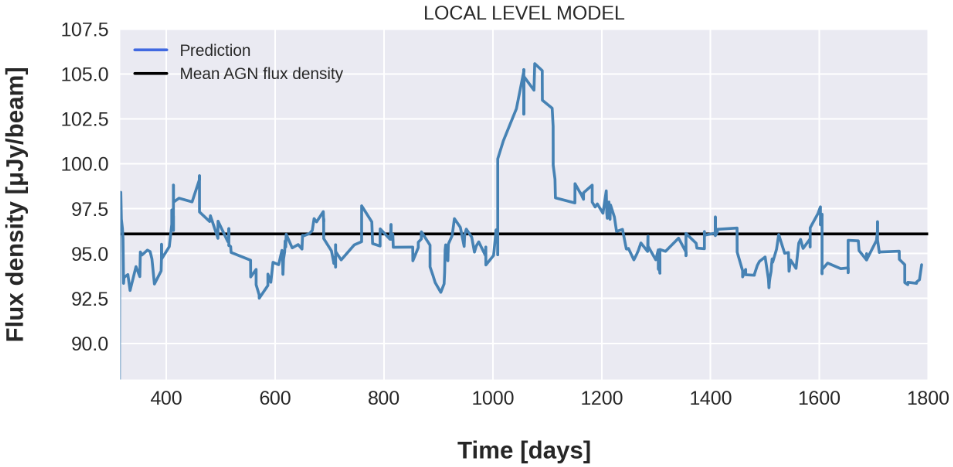}
		\caption{ \small {\label{LLM_trend} {Local Level Model compared to the mean flux density. The blue trend is the prediction of the Local Level Model while the black line is the level of the mean flux density of the light curve.  }}}
\end{figure*}

To detect a transient within the light curve in Fig.~\ref{GW_AGN2} it is necessary to analyse the trend of the time series which is successfully reproduced by the Local Level Model. If there is a detectable transient within the light curve, there must be a discrepancy between the time series model trend and the mean flux density value of the data. In Fig.~\ref{LLM_trend} we can see that there is a large gap between 1000 and 1200 days of observations. The trend model is mostly around the mean value of 96.1 $\mu$Jy with a gap between the trend and the mean flux density below 4 except for this time interval. The maximum gradient between the trend and the mean flux density is 9.47 after 1077 days of observations.

\subsection{Change points localisation}
A method for detecting a transient within an active galaxy was proposed above. The light curve interval where the transient has been detected is approximately between 1000 and 1200 days. In this interval the flux density must be dominated by the transient source. This leads to the problem of understanding when the transient starts dominating, when it ends dominating and how long the burst dominates. In other words, we know that this interval is between 1000 and 1200 days but this is a rough indication. We do not know the exact time interval dominated by the transient. The solution is to detect change points in the time series $trend$. 

\begin{figure}
        \hspace*{0.3cm} 
		\includegraphics[scale=0.23]{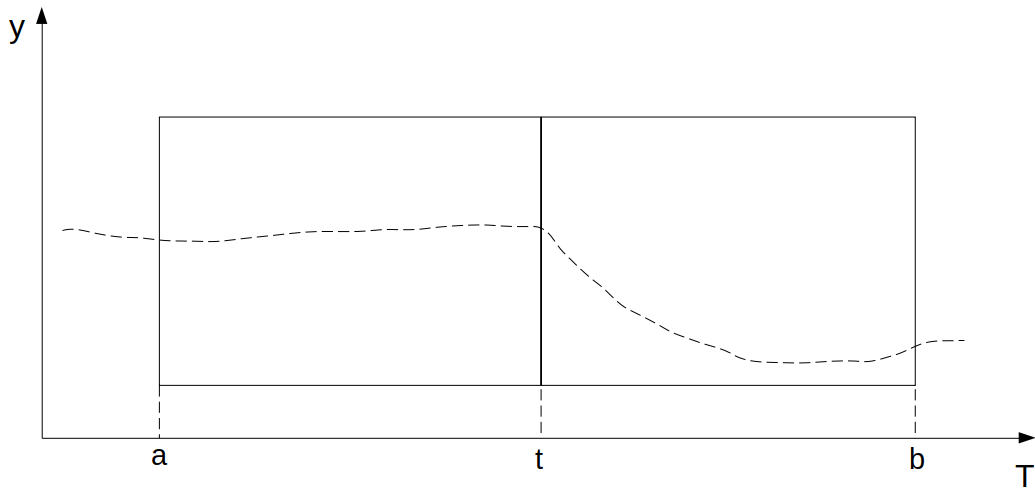}
		\caption{ \small {\label{slid_windows} {Illustration of a change point at the time $T = t$ and two adjacent sliding windows in a given time series. The dashed line is the signal. The two rectangles are two sliding windows covering a time interval $a \leq t$ and $b \geq t$, respectively.   }}}
\end{figure}

A change point divides a time series in two subsets where each subset has its own statistical characteristics such as mean and variance \citep{Sharma2016}. The statistical properties of each subset are different from the statistical properties of the other subset. In Fig.~\ref{slid_windows} there is graph showing a change point example. 

To detect change points in the AGN light curve a Window-based change point detection method was applied. This method is based on the usage of sliding windows. A sliding window is a subset of $n$ data points in the main time series. The Window-based change point detection method consists of computing the discrepancy between two adjacent windows sliding along the signal $y$ \citep{Tru2020}. The graph in Fig.~\ref{slid_windows} shows a time series divided in two adjacent windows. The discrepancy between two sliding windows is given by the following formula: 

\begin{equation}  \label{eq.1_4}
\centering
d(y_{a...t, b...t}) = c(y_{a...t}) - c(y_{a...b}) - c(y_{b...t}),
\end{equation}

\noindent where $d$ is the discrepancy, $y$ is the signal, $a$, $b$ and $t$ define the time intervals covered by the two windows in Fig.~\ref{slid_windows}. The function $c()$ is called cost function and is used to determine a difference of the statistical properties of the two subsets (windows) of the time series $T$. The discrepancy has its largest values when is calculated between two dissimilar subsets.

To detect the change points in the AGN light curve we assumed the following conditions:

\begin{itemize}
    \item the number of change points in the time series is 2 as we expect the GW source to have a burst behaviour with a rising peak and then a decline;
    \item the burst is approximately located between 1000 and 1200 days as shown in Fig.~\ref{LLM_trend}.
    
\end{itemize}

The cost function adopted is the least squared deviation defined by the following equation:

\begin{equation}  \label{eq.1_4}
\centering
     c(y_{I}) = \sum\limits_{t=1}^N ||y_{t} - \bar{y}||^{2},
\end{equation}

\noindent where $\bar{y}$ is the mean of the signal $y$ in the interval $I$. This function can be used to detect mean-shift in a signal and is implemented by the $Python$ package $ruptures$.

\begin{figure*}
		\includegraphics[scale=0.53]{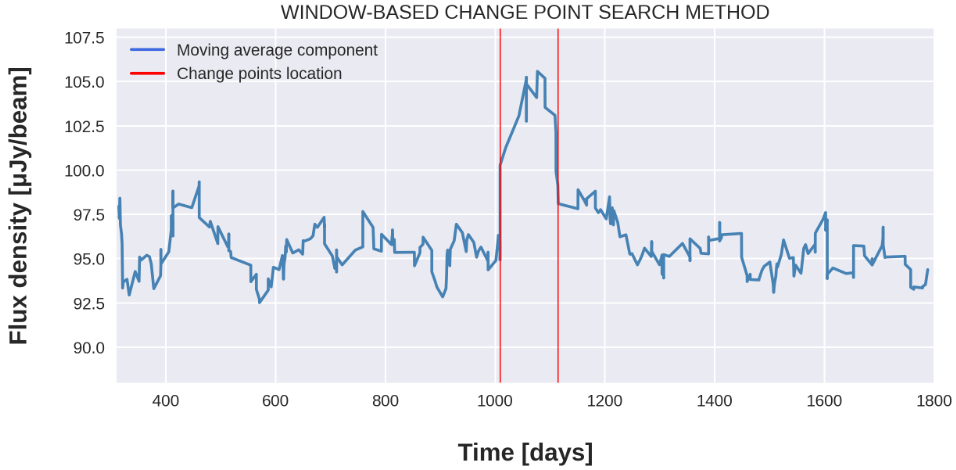}
		\caption{ \small {\label{change_points} {Change points location detected with the window-based method.  }}}
\end{figure*}

The length of the sliding windows was chosen with an empirical method to detect change points in an interval consistent with the one described by the Local Level Model in Fig~\ref{LLM_trend} (1000 - 1200 days). Setting for each window a number of samples (data points) 12 $\leq$ n $\leq$ 13  gave change points at $t = 1043$ days and $t = 1109$ days. With 14 $\leq$ n $\leq$ 39  we got change points at $t = 1043$ days and $t = 1114.2$ days. Using the interval 63 $\leq$ n $\leq$ 68 gave change points at $t = 1009$ days and $t = 1188$ days. Values of $n$ outside of the intervals just mentioned gave no change points within the interval 1000-1200 days. We also decided not to consider the change point at $t = 1188$ days as it is far off the transient peak activity in the light curve. We chose the widest possible region based on the remaining change points to make sure the transient activity was fully included. Thus, we established that the most suitable change points are at $t = 1009$ days and $t = 1114.2$ days. In Fig.~\ref{change_points} the location of the change points is represented.

The detection of the change points gives the precise segment of the light curve which is dominated by the gravitational waves event. The light curve subset outside this segment is instead dominated by the AGN. 

The usage of the Local Level Model combined with the change point detection hence shows that the transient source dominated between 1009 and 1114 days. Note that this estimation is close to the actual one. Because we used a simulated AGN, we actually know which data points are originated by the GW event and which data points are simulated. A check of the origin of every single data point revealed that the GW dominated between 1001 and 1091 days which is not very far from the interval predicted by the change points detection method. 
The first change point differs from the predicted one by 8 days while the second change point differs by 23 days. This may seem a large gap. However, the whole light curve covers a much longer time interval which is roughly 350-1800 days corresponding to 1450 days in total.

\section{Summary and conclusions}\label{conclusions}

In this work, innovative methods to analyse the variability of time series in Astrophysics were described and applied to the transient GW170817 light curve. The current metrics involving $\chi^{2}$ and modulation index used in Astronomy only provide an overall description of the variability. They do not give any description regarding the order of the data points. This is instead possible by using State Space Models. In detail, this study showed that it is possible to examine variability with several approaches. These methods and their possible developments are outlined below:

\begin{itemize}
    \item Time series decomposition into components. An example is given by the SSARIMA(1,2,1) model used to fit the GW light curve. A moving average component describes a stochastic behaviour. A simple Local Level Model also provides a description about the trend of the time series between two data points thanks to the level component. This tells us if there is a decline, a rise or a flat trend in the time series for a given time $t$.  

    \item An example of stochastic behaviour description was given with the model SSARIMA(3,1,25). We showed an alternative way for describing variability. The moving average component order MA = 25 describes a strong stochastic behaviour. This property would not be described by the traditional metrics such as modulation index and $\chi^{2}$.

    \item Time series stationarity. Testing time series stationarity provides other information such as a constant or not constant mean of the time series.

    \item State Space Models could allow the detection of transients even if hosted by another source such as a variable active galaxy. This is the same scenario of the fast radio burst FRB 150418 \citep{Kea2016}. State Space Models could indeed be used to detect bursts within an active galaxy and establish if a transient is detected or not. We reproduced a scenario of gravitational waves source hosted by a variable galaxy.  
    
    \item State Space Models may also open the possibility of classifying different transient and variable sources. The Local Level Model successfully fits the gravitational wave event GW170817. This model is suitable for a sheer burst with a fast rise and decline. Other transients and variables with a different behaviour may need a different model. Hence for each transient class we may have a different model. State Space Models combined with supervised machine learning techniques of regression analysis \citep{Shu_Liu2021} could be used for classifying transients. 

    \item The usage of State Space Models also opens new scenarios such as the usage of change points detection in time series. Thanks to change point detection methods it is possible to know the exact locations of a transient activity in a light curve such as in the scenario of the GW event hosted by an active galaxy. Furthermore, change points detection could be used even for detecting the given time $t$ when a light curve shows a variable behaviour.

\end{itemize}

\noindent Note that we did not use time series models on upper-limits measurements. However, State Space Models can accurately capture upper-limits, just as they handle proper detections. In the case of upper-limits, we only need the model confidence region encompassing values below the upper limit whilst values above it should be disregarded. This approach may work if the upper limit is far away from the other flux density measurements. Otherwise, we may need to adopt Non-Gaussian State Space Models \citep{non-gauss-mod}. Thus, the usage of these models would be a further research development for upper limits.

\section*{Acknowledgements}
This research
was supported by the Australian Government Research Training Program and
the Australian Research Council (ARC). We thank Dougal Dobie who kindly provided us the flux density measurements of the gravitional waves source GW170817 (\citealt{Dob2018}; \citealt{Dob2019}). We also thank the Reviewers for their worth comments which improved the research quality of this work.

The Australian SKA Pathfinder is part of the Australia Telescope National Facility which is managed by CSIRO. Operation of ASKAP is funded by the Australian Government with support from the National Collaborative Research Infrastructure Strategy. ASKAP uses the resources of the Pawsey Supercomputing Centre. Establishment of ASKAP, the Murchison Radio-astronomy Observatory and the Pawsey Supercomputing Centre are initiatives of the Australian Government, with support from the Government of Western Australia and the Science and Industry Endowment Fund. We acknowledge the Wajarri Yamatji people as the traditional owners of the Observatory site.

\appendix

\section{State Space Models in Python}

In this Appendix, fragments of the Python code used to fit State Space Models are shown.  

The code shows how to fit different State Space Models on time series and how to estimate statistical parameters (AIC, BIC, HQIC and Heteroskedasticity).

\begin{figure*}
        \hspace*{2.8cm} 
		\includegraphics[scale=0.25]{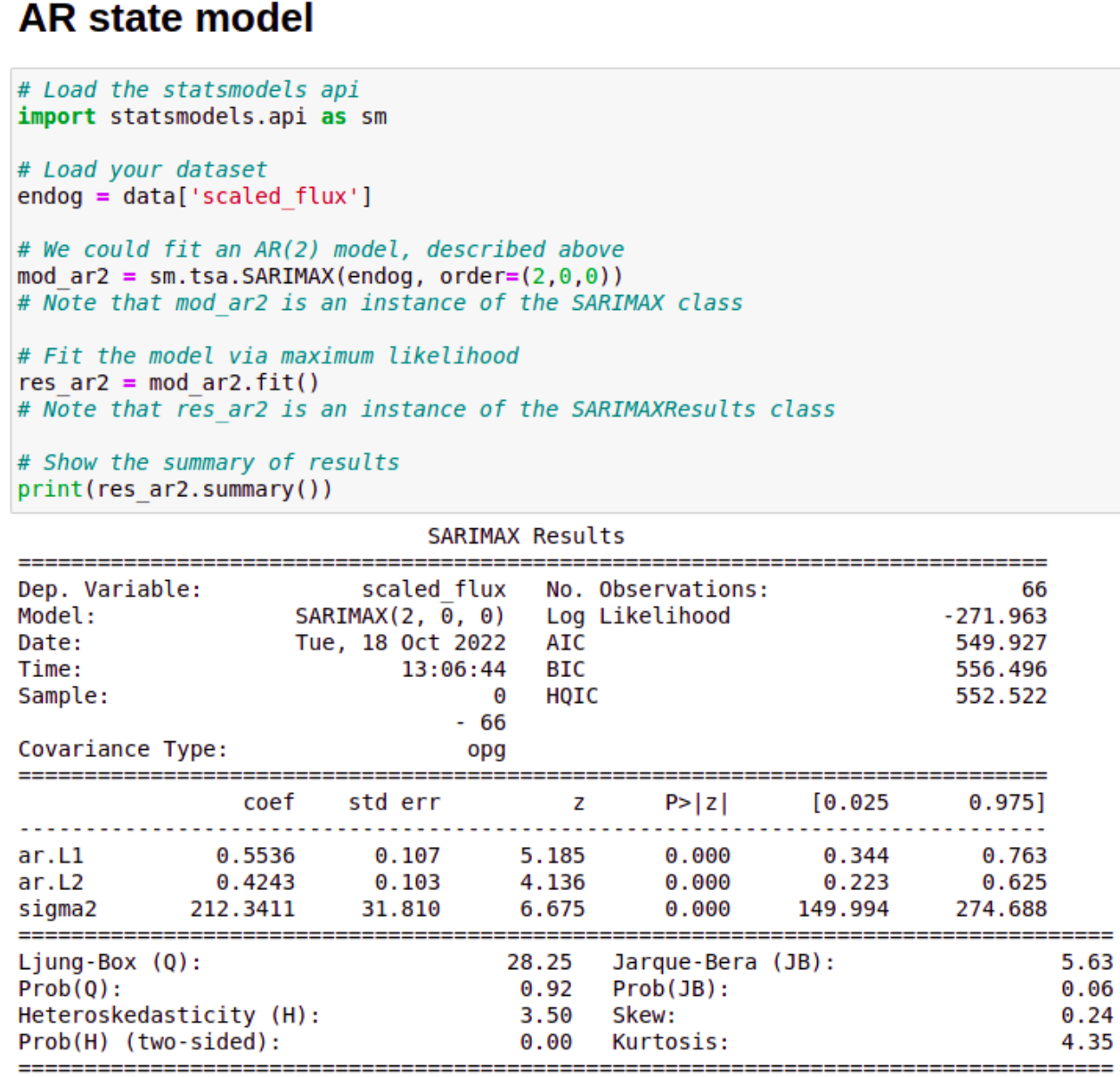}
		\caption{ \small {\label{appendix_AR} {State Space Autoregression Model analysis.  }}}
\end{figure*}

The models adopted are based on the Python library $statsmodels$ which contains a list of functions specifically designed for using State Space Models. In Fig.~\ref{appendix_AR} we can see the code used for the statistical analysis of the State Space Autoregression Model. The model is defined by the function SARIMAX() of the $stasmodels$ library. The SARIMAX() function is used for State Space ARIMA Models and in this case an ARIMA model of order (p=2, d=0, q=0) is used. This model can be seen as an Autoregression process of order 2. In Fig.~\ref{appendix_AR} we can see that several parameters are estimated such as AIC, BIC, HQIC and Heteroskedasticity. The usage of the other parameters was beyond the goals of this work but it is possible to learn more thanks to the $statsmodels$ documentation\footnote{\url{https://www.statsmodels.org/stable/index.html}}.

\newpage

\begin{figure*}
       \hspace*{2.0cm} 
		\includegraphics[scale=0.25]{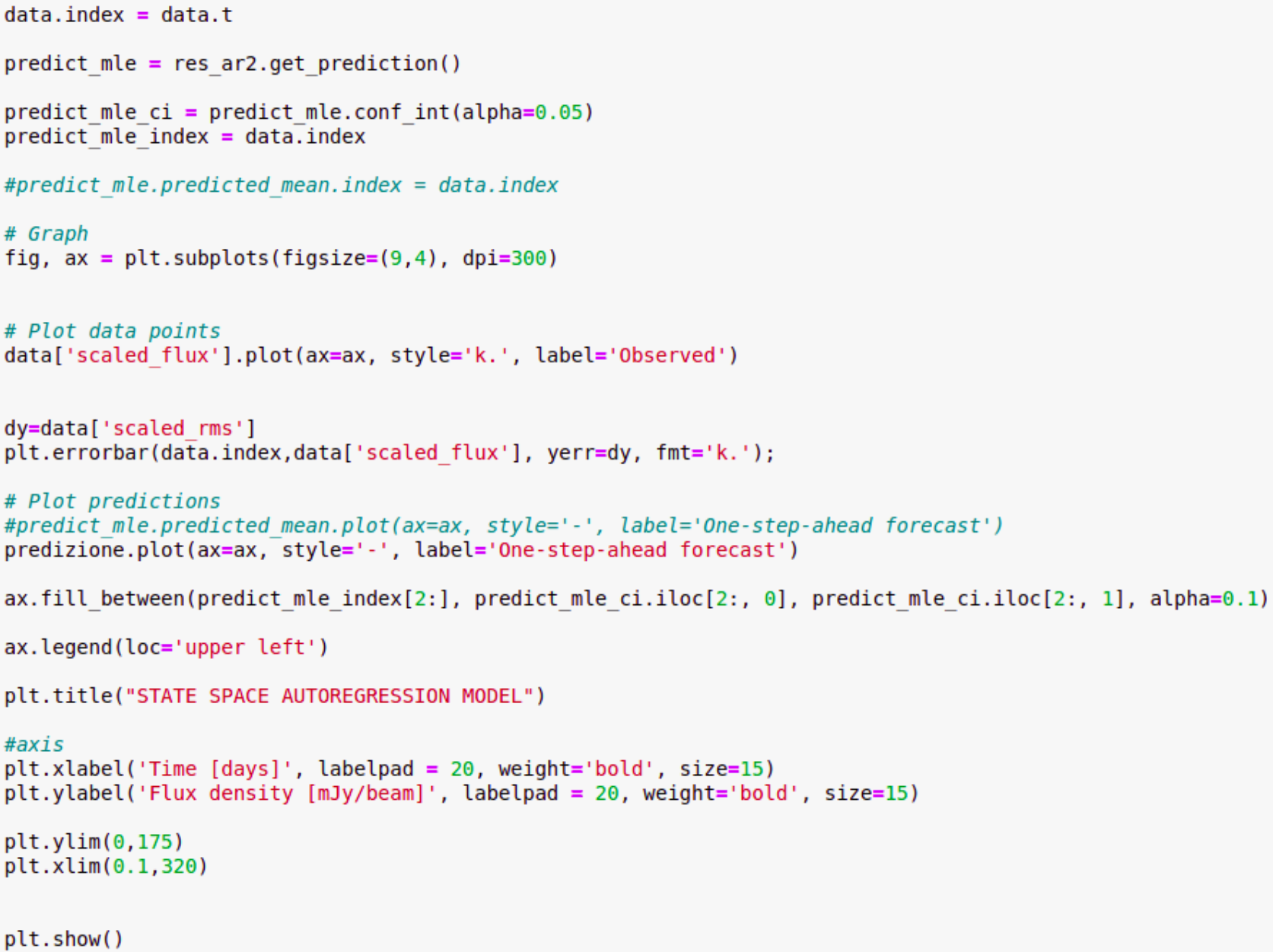}
		\caption{ \small {\label{appendix_AR2} {Code for fitting State Space Autoregression Model on a time series.  }}}
\end{figure*}

The code for fitting a time series with the Autoregression process is in Fig.~\ref{appendix_AR2}. The code also includes the confidence region of the model. 

\clearpage

\newpage

\begin{figure*}
        \hspace*{2.8cm} 
		\includegraphics[scale=0.25]{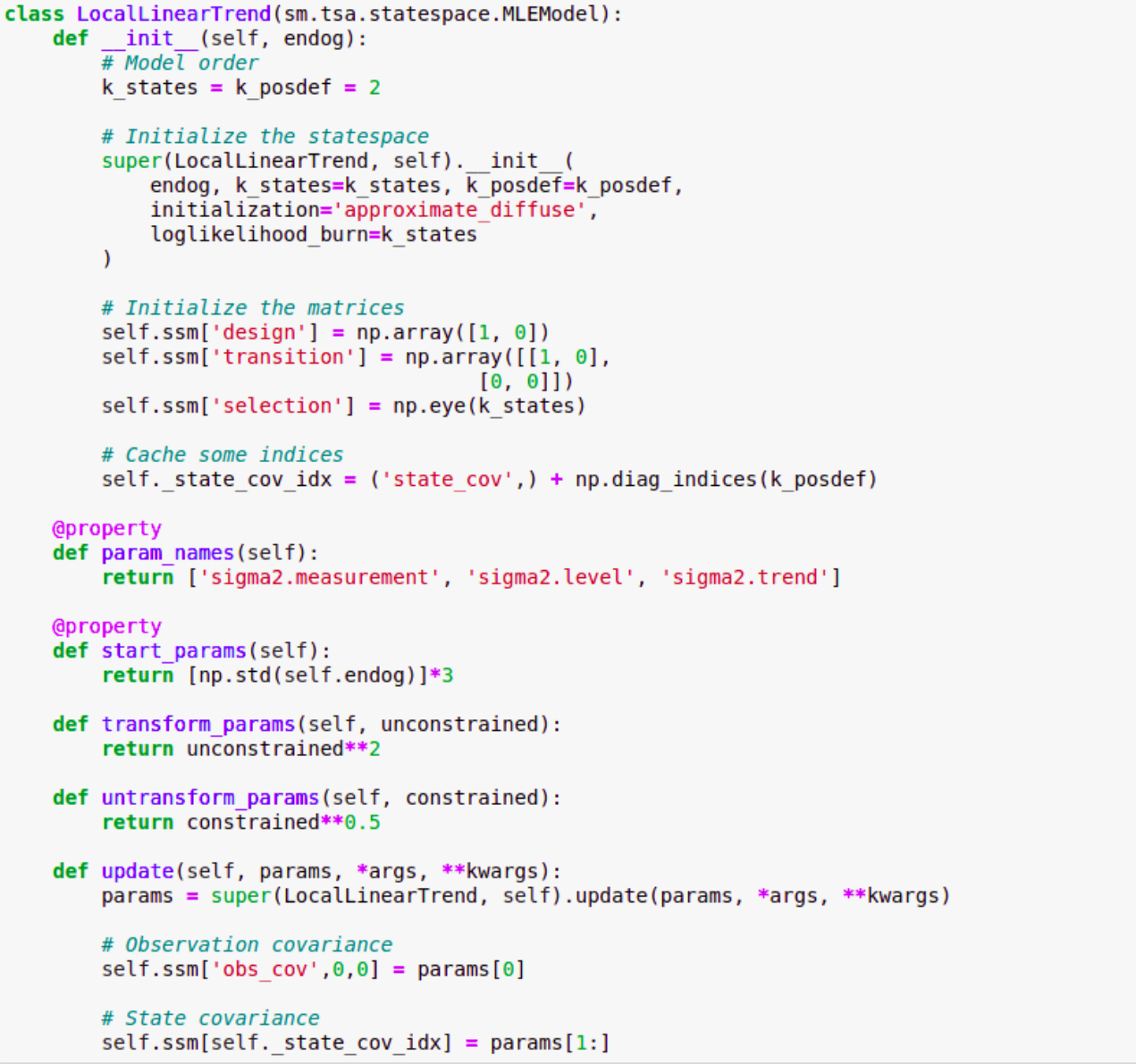}
		\caption{ \small {\label{appendix_LLM} {Code for defining the LocalLinearTrend class which is used for Local Liner Trend Models and Local Level Models.  }}}
\end{figure*}

\begin{figure*}
        \hspace*{5cm} 
		\includegraphics[scale=0.53]{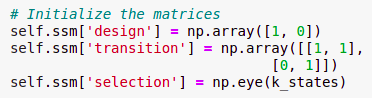}
		\caption{ \small {\label{appendix_LTM} {Matrices defining the Local Linear Trend Model.  }}}
\end{figure*}

\begin{figure*}
       \hspace*{1.8cm} 
		\includegraphics[scale=0.25]{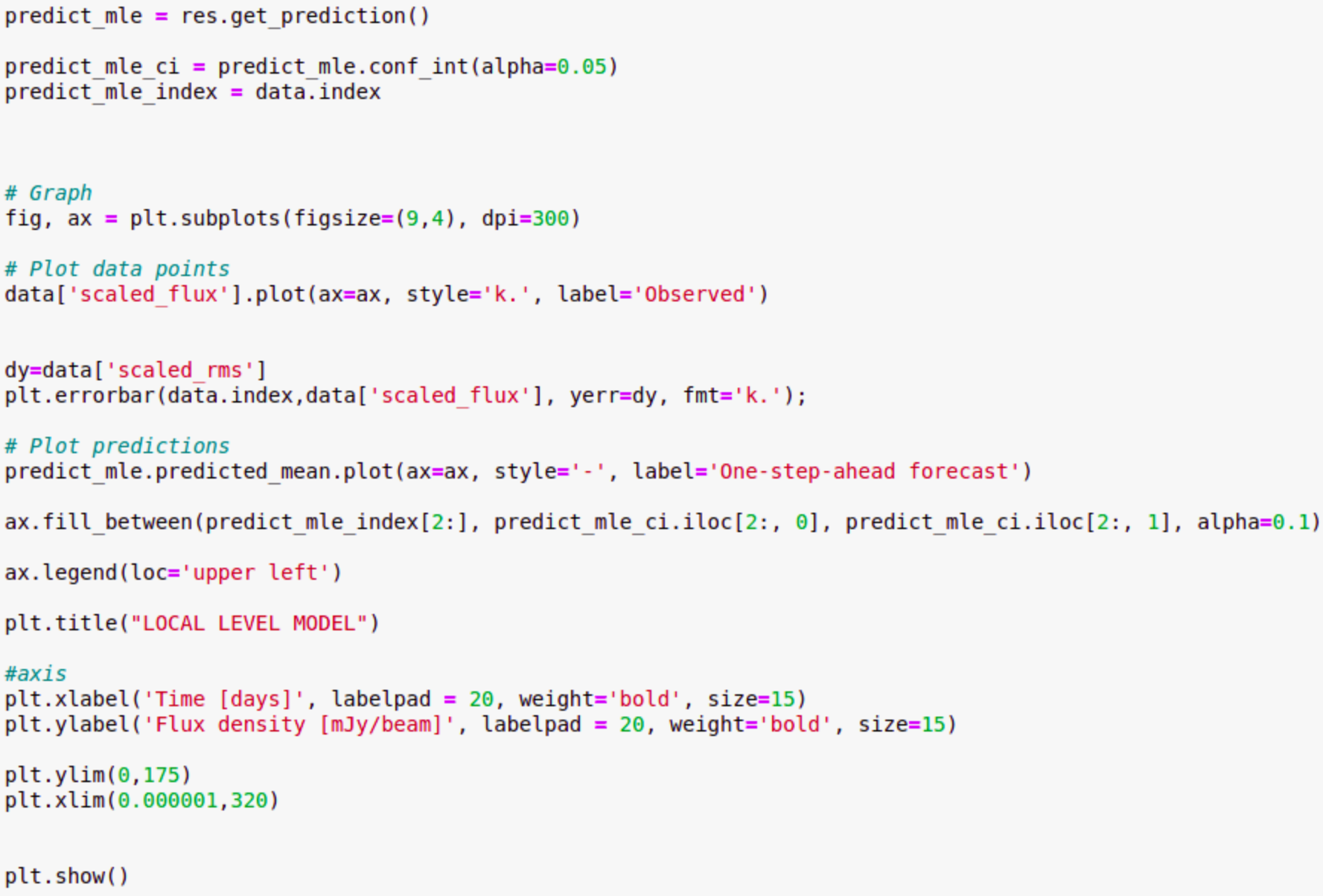}
		\caption{ \small {\label{appendix_LLM3} {Code for fitting the Local Level Model on a time series.  }}}
\end{figure*}

In Fig.~\ref{appendix_LLM} there is the code adopted to define the Local Level Model. This code is based on the LocalLinearTrend class. As the name suggests this class can be used to also define the Local Linear Trend Model. The Local Linear Trend Model is a generalisation of the Local Level Model. The difference between the two models is a slope component in the Local Linear Trend Model not included in the Local Level Model. In terms of coding, this means to have different coefficients in the matrices defining the model (compare the matrices in Fig.~\ref{appendix_LLM} and \ref{appendix_LTM}). Note also Fig.~\ref{appendix_LLM3} where there is the code used for plotting the Local Level Model on a time series.

\clearpage

\bibliographystyle{mnras}
\bibliography{biblio}

\begin{thebibliography}{}
\makeatletter
\relax
\def\mn@urlcharsother{\let\do\@makeother \do\$\do\&\do\#\do\^\do\_\do\%\do\~}
\def\mn@doi{\begingroup\mn@urlcharsother \@ifnextchar [ {\mn@doi@} {\mn@doi@[]}}
\def\mn@doi@[#1]#2{\def\@tempa{#1}\ifx\@tempa\@empty \href {http://dx.doi.org/#2} {doi:#2}\else \href {http://dx.doi.org/#2} {#1}\fi \endgroup}
\def\mn@eprint#1#2{\mn@eprint@#1:#2::\@nil}
\def\mn@eprint@arXiv#1{\href {http://arxiv.org/abs/#1} {{\tt arXiv:#1}}}
\def\mn@eprint@dblp#1{\href {http://dblp.uni-trier.de/rec/bibtex/#1.xml} {dblp:#1}}
\def\mn@eprint@#1:#2:#3:#4\@nil{\def\@tempa {#1}\def\@tempb {#2}\def\@tempc {#3}\ifx \@tempc \@empty \let \@tempc \@tempb \let \@tempb \@tempa \fi \ifx \@tempb \@empty \def\@tempb {arXiv}\fi \@ifundefined {mn@eprint@\@tempb}{\@tempb:\@tempc}{\expandafter \expandafter \csname mn@eprint@\@tempb\endcsname \expandafter{\@tempc}}}

\bibitem[\protect\citeauthoryear{A.~Richard~Thompson}{A.~Richard~Thompson}{2016}]{astro_bible}
A.~Richard~Thompson James M.~Moran G. W. S.~J.,  2016, Interferometry and Synthesis in Radio Astronomy.
Springer

\bibitem[\protect\citeauthoryear{Abbott et~al.,}{Abbott et~al.}{2016}]{Abb2016}
Abbott B.~P.,  et~al., 2016, \mn@doi [Phys. Rev. Lett.] {10.1103/PhysRevLett.116.061102}, 116, 061102

\bibitem[\protect\citeauthoryear{Abbott et~al.,}{Abbott et~al.}{2017}]{Abb2017}
Abbott B.,  et~al., 2017, \mn@doi [Physical Review Letters] {10.1103/physrevlett.119.161101}, 119

\bibitem[\protect\citeauthoryear{{Alexander}, {Fong}  \& {Berger}}{{Alexander} et~al.}{2017}]{Ale2017}
{Alexander} K.~D.,  {Fong} W.,   {Berger} E.,  2017, GRB Coordinates Network, \href {https://ui.adsabs.harvard.edu/abs/2017GCN.21545....1A} {21545, 1}

\bibitem[\protect\citeauthoryear{Alexander et~al.,}{Alexander et~al.}{2018}]{Alexander2018}
Alexander K.~D.,  et~al., 2018, \mn@doi [The Astrophysical Journal Letters] {10.3847/2041-8213/aad637}, 863, L18

\bibitem[\protect\citeauthoryear{{Bannister}, {Lynch}, {Kaplan}, {Murphy}, {Dobie}  \& {VAST Collaboration.}}{{Bannister} et~al.}{2017a}]{Ban2017}
{Bannister} K.,  {Lynch} C.,  {Kaplan} D.,  {Murphy} T.,  {Dobie} D.,   {VAST Collaboration.} 2017a, GRB Coordinates Network, \href {https://ui.adsabs.harvard.edu/abs/2017GCN.21559....1B} {21559, 1}

\bibitem[\protect\citeauthoryear{{Bannister}, {Shannon}, {Hotan}, {James}, {Macquart}, {Oslowski}, {Farah}  \& {Askap Collaboration.}}{{Bannister} et~al.}{2017b}]{Ban2017b}
{Bannister} K.,  {Shannon} R.,  {Hotan} A.,  {James} C.,  {Macquart} J.~P.,  {Oslowski} S.,  {Farah} W.,   {Askap Collaboration.} 2017b, GRB Coordinates Network, \href {https://ui.adsabs.harvard.edu/abs/2017GCN.21562....1B} {21562, 1}

\bibitem[\protect\citeauthoryear{{Bannister}, {Shannon}, {Hotan}, {James}, {Oslowski}  \& {Farah}}{{Bannister} et~al.}{2017c}]{Ban2017c}
{Bannister} K.,  {Shannon} R.,  {Hotan} A.,  {James} C.,  {Oslowski} S.,   {Farah} W.,  2017c, GRB Coordinates Network, \href {https://ui.adsabs.harvard.edu/abs/2017GCN.21671....1B} {21671, 1}

\bibitem[\protect\citeauthoryear{Barreto \& Howland}{Barreto \& Howland}{2006}]{Bar2006}
Barreto H.,  Howland F.,  2006, Introductory econometrics. Using Monte Carlo simulation with Microsoft Excel. With CD-ROM.
Cambridge University Press, \url {https://www.researchgate.net/publication/264950115_Introductory_econometrics_Using_Monte_Carlo_simulation_with_Microsoft_Excel_With_CD-ROM}

\bibitem[\protect\citeauthoryear{Bell et~al.,}{Bell et~al.}{2011}]{Bell2011}
Bell M.~E.,  et~al., 2011, \mn@doi [Monthly Notices of the Royal Astronomical Society] {10.1111/j.1365-2966.2010.17692.x}, 411, 402

\bibitem[\protect\citeauthoryear{Bell, Huynh, Hancock, Murphy, Gaensler, Burlon, Trott  \& Bannister}{Bell et~al.}{2015}]{Bell2015}
Bell M.~E.,  Huynh M.~T.,  Hancock P.,  Murphy T.,  Gaensler B.~M.,  Burlon D.,  Trott C.,   Bannister K.,  2015, \mn@doi [Monthly Notices of the Royal Astronomical Society] {10.1093/mnras/stv882}, 450, 4221

\bibitem[\protect\citeauthoryear{Berger}{Berger}{2014}]{Ber2014}
Berger E.,  2014, \mn@doi [Annual Review of Astronomy and Astrophysics] {10.1146/annurev-astro-081913-035926}, 52, 43

\bibitem[\protect\citeauthoryear{Bhattacharyya, Ghosh, Chatterjee  \& Das}{Bhattacharyya et~al.}{2020}]{Bhat2020}
Bhattacharyya S.,  Ghosh R.,  Chatterjee R.,   Das N.,  2020, \mn@doi [The Astrophysical Journal] {10.3847/1538-4357/ab91a8}, 897, 25

\bibitem[\protect\citeauthoryear{Boone}{Boone}{2019}]{Boo2019}
Boone K.,  2019, \mn@doi [The Astronomical Journal] {10.3847/1538-3881/ab5182}, 158, 257

\bibitem[\protect\citeauthoryear{Bramich}{Bramich}{2008}]{Bra2008}
Bramich D.~M.,  2008, \mn@doi [Monthly Notices of the Royal Astronomical Society: Letters] {10.1111/j.1745-3933.2008.00464.x}, 386, L77

\bibitem[\protect\citeauthoryear{Brockwell \& Davis}{Brockwell \& Davis}{2010}]{Broc2010}
Brockwell Davis 2010, Introduction to Time Series and Forecasting.
Springer

\bibitem[\protect\citeauthoryear{Brooks}{Brooks}{2008}]{Brook2008}
Brooks C.,  2008, Univariate time series modelling and forecasting.
Cambridge University Press, \url {https://doi.org/10.1017/CBO9780511841644.006}

\bibitem[\protect\citeauthoryear{Buckland, Newman, Thomas  \& Koesters}{Buckland et~al.}{2004}]{BUCK2004}
Buckland S.,  Newman K.,  Thomas L.,   Koesters N.,  2004, \mn@doi [Ecological Modelling] {https://doi.org/10.1016/j.ecolmodel.2003.08.002}, 171, 157

\bibitem[\protect\citeauthoryear{Burnham \& Anderson}{Burnham \& Anderson}{2002}]{Bur2002}
Burnham K.~P.,  Anderson D.~R.,  2002, Model Selection and Multimodel Inference.
Springer, \url {http://www.elcom-hu.com/Mshtrk/Statstics/9th\%20txt\%20book.pdf}

\bibitem[\protect\citeauthoryear{Burnham \& Anderson}{Burnham \& Anderson}{2004}]{Ken2004}
Burnham K.~P.,  Anderson D.~R.,  2004, \mn@doi [Sociological Methods \& Research] {10.1177/0049124104268644}, 33, 261

\bibitem[\protect\citeauthoryear{Camilo}{Camilo}{2018}]{Meerkat}
Camilo F.,  2018.
Nat. Astron., 2, 594

\bibitem[\protect\citeauthoryear{Dobie et~al.,}{Dobie et~al.}{2018}]{Dob2018}
Dobie D.,  et~al., 2018, \mn@doi [The Astrophysical Journal] {10.3847/2041-8213/aac105}, 858, L15

\bibitem[\protect\citeauthoryear{Dobie, Murphy, Kaplan, Ghosh, Bannister  \& Hunstead}{Dobie et~al.}{2019}]{Dob2019}
Dobie D.,  Murphy T.,  Kaplan D.~L.,  Ghosh S.,  Bannister K.~W.,   Hunstead R.~W.,  2019, \mn@doi [Publications of the Astronomical Society of Australia] {10.1017/pasa.2019.9}, 36, e019

\bibitem[\protect\citeauthoryear{Driessen et~al.,}{Driessen et~al.}{2019}]{Dri2020}
Driessen L.~N.,  et~al., 2019, \mn@doi [Monthly Notices of the Royal Astronomical Society] {10.1093/mnras/stz3027}, 491, 560

\bibitem[\protect\citeauthoryear{Durbin \& Koopman}{Durbin \& Koopman}{2012}]{Dur2012}
Durbin J.,  Koopman S.~J.,  2012, Time Series Analysis by State Space Methods.
Oxford University Press, Incorporated, \url {https://www.researchgate.net/publication/227468262_Time_Series_Analysis_by_State_Space_Methods}

\bibitem[\protect\citeauthoryear{Du Toit, Grobler  \& Ludick}{Du Toit et~al.}{2024}]{Du2024}
Du Toit C.~D.,  Grobler T.~L.,   Ludick D.~J.,  2024, \mn@doi [Monthly Notices of the Royal Astronomical Society] {10.1093/mnras/stae892}, 530, 613

\bibitem[\protect\citeauthoryear{Feigelson, Babu  \& Caceres}{Feigelson et~al.}{2018}]{Fei2018}
Feigelson E.~D.,  Babu G.~J.,   Caceres G.~A.,  2018, \mn@doi [Frontiers in Physics] {10.3389/fphy.2018.00080}, 6

\bibitem[\protect\citeauthoryear{Fender, Anderson, Osten, Staley, Rumsey, Grainge  \& Saunders}{Fender et~al.}{2014}]{Fen2014}
Fender R.~P.,  Anderson G.~E.,  Osten R.,  Staley T.,  Rumsey C.,  Grainge K.,   Saunders R. D.~E.,  2014, \mn@doi [Monthly Notices of the Royal Astronomical Society: Letters] {10.1093/mnrasl/slu165}, 446, L66

\bibitem[\protect\citeauthoryear{Gabbiani \& Cox}{Gabbiani \& Cox}{2017}]{Gab2017}
Gabbiani F.,  Cox S.~J.,  2017, Mathematics for Neuroscientists.
Elsevier

\bibitem[\protect\citeauthoryear{Gardner et~al.,}{Gardner et~al.}{2006}]{Gar2006}
Gardner J.~P.,  et~al., 2006, \mn@doi [Space Science Reviews] {10.1007/s11214-006-8315-7}, 123, 485–606

\bibitem[\protect\citeauthoryear{{Giroletti, M.}, {Marcote, B.}, {Garrett, M. A.}, {Paragi, Z.}, {Yang, J.}, {Hada, K.}, {Muxlow, T. W. B.}  \& {Cheung, C. C.}}{{Giroletti, M.} et~al.}{2016}]{Giro2016}
{Giroletti, M.} {Marcote, B.} {Garrett, M. A.} {Paragi, Z.} {Yang, J.} {Hada, K.} {Muxlow, T. W. B.}  {Cheung, C. C.} 2016, \mn@doi [A&A] {10.1051/0004-6361/201629172}, 593, L16

\bibitem[\protect\citeauthoryear{{Gupta}, {Muthukrishna}  \& {Lochner}}{{Gupta} et~al.}{2024}]{Gup2024}
{Gupta} R.,  {Muthukrishna} D.,   {Lochner} M.,  2024, \mn@doi [arXiv e-prints] {10.48550/arXiv.2403.14742}, \href {https://ui.adsabs.harvard.edu/abs/2024arXiv240314742G} {p. arXiv:2403.14742}

\bibitem[\protect\citeauthoryear{{Hallinan} et~al.,}{{Hallinan} et~al.}{2017}]{Hal2017}
{Hallinan} G.,  et~al., 2017, \mn@doi [Science] {10.1126/science.aap9855}, \href {https://ui.adsabs.harvard.edu/abs/2017Sci...358.1579H} {358, 1579}

\bibitem[\protect\citeauthoryear{Hamilton}{Hamilton}{1994}]{HAM1994}
Hamilton J.~D.,  1994, in , Vol.~4, Handbook of Econometrics.
Elsevier, pp 3039--3080, \mn@doi{https://doi.org/10.1016/S1573-4412(05)80019-4}, \url {https://www.sciencedirect.com/science/article/pii/S1573441205800194}

\bibitem[\protect\citeauthoryear{{Hern{\'a}ndez-Afonso} \& {Baena-Gall{\'e}}}{{Hern{\'a}ndez-Afonso} \& {Baena-Gall{\'e}}}{2023}]{Her2023}
{Hern{\'a}ndez-Afonso} J.,  {Baena-Gall{\'e}} R.,  2023, in {Manteiga} M.,  {Bellot} L.,  {Benavidez} P.,  {de Lorenzo-C{\'a}ceres} A.,  {Fuente} M.~A.,  {Mart{\'\i}nez} M.~J.,  {V{\'a}zquez Acosta} M.,   {Dafonte} C.,  eds, Highlights on Spanish Astrophysics XI. p.~393

\bibitem[\protect\citeauthoryear{Hewish, Bell, Pilkington, Scott  \& Collins}{Hewish et~al.}{1968}]{Hew1968}
Hewish A.,  Bell S.~J.,  Pilkington J.~D.~H.,  Scott P.~F.,   Collins R.~A.,  1968, Observation of a Rapidly Pulsating Radio Source.
Nature

\bibitem[\protect\citeauthoryear{Hotan et~al.,}{Hotan et~al.}{2021}]{ASKAP}
Hotan A.~W.,  et~al., 2021, \mn@doi [Publications of the Astronomical Society of Australia] {10.1017/pasa.2021.1}, 38, e009

\bibitem[\protect\citeauthoryear{J.~Durbin}{J.~Durbin}{2000}]{non-gauss-mod}
J.~Durbin S. J.~K.,  2000, Time Series Analysis of Non-Gaussian Observations Based on State Space Models from Both Classical and Bayesian Perspectives.
Oxford University Press

\bibitem[\protect\citeauthoryear{Jiménez}{Jiménez}{2021}]{jiménez2021}
Jiménez J.~C.,  2021, BayesDLMfMRI: Bayesian Matrix-Variate Dynamic Linear Models for Task-based fRMI Modeling in R (\mn@eprint {arXiv} {2111.01318})

\bibitem[\protect\citeauthoryear{Johnston et~al.,}{Johnston et~al.}{2016}]{Joh2016}
Johnston S.,  et~al., 2016, \mn@doi [Monthly Notices of the Royal Astronomical Society] {10.1093/mnras/stw2808}, 465, 2143

\bibitem[\protect\citeauthoryear{Karpenka, Feroz  \& Hobson}{Karpenka et~al.}{2012}]{Kar2013}
Karpenka N.~V.,  Feroz F.,   Hobson M.~P.,  2012, \mn@doi [Monthly Notices of the Royal Astronomical Society] {10.1093/mnras/sts412}, 429, 1278

\bibitem[\protect\citeauthoryear{Keane et~al.,}{Keane et~al.}{2016}]{Kea2016}
Keane E.~F.,  et~al., 2016, The host galaxy of a fast radio burst.
Nature, \url {https://ui.adsabs.harvard.edu/abs/2016Natur.530..453K}

\bibitem[\protect\citeauthoryear{Kelly, Becker, Sobolewska, Siemiginowska  \& Uttley}{Kelly et~al.}{2014}]{Kel2014}
Kelly B.~C.,  Becker A.~C.,  Sobolewska M.,  Siemiginowska A.,   Uttley P.,  2014, \mn@doi [The Astrophysical Journal] {10.1088/0004-637x/788/1/33}, 788, 33

\bibitem[\protect\citeauthoryear{Koller \& Friedman}{Koller \& Friedman}{2009}]{koller2009}
Koller D.,  Friedman N.,  2009, Probabilistic graphical models: principles and techniques.
MIT press

\bibitem[\protect\citeauthoryear{{Konig} \& {Timmer}}{{Konig} \& {Timmer}}{1997}]{Kon1997}
{Konig} M.,  {Timmer} J.,  1997, Analyzing X-ray variability by linear state space models.
Springer

\bibitem[\protect\citeauthoryear{Koopman \& Durbin}{Koopman \& Durbin}{2012}]{Koop2012}
Koopman S.~J.,  Durbin J.,  2012, Time Series Analysis by State Space Methods: Second Edition.
OUP Catalogue, Oxford University Press

\bibitem[\protect\citeauthoryear{Kwiatkowski, Phillips, Schmidt  \& Shin}{Kwiatkowski et~al.}{1992}]{Kwi1992}
Kwiatkowski D.,  Phillips P.~C.,  Schmidt P.,   Shin Y.,  1992, \mn@doi [Journal of Econometrics] {https://doi.org/10.1016/0304-4076(92)90104-Y}, 54, 159

\bibitem[\protect\citeauthoryear{Lazio, Waltman, Ghigo, Fiedler, Foster  \& Johnston}{Lazio et~al.}{2001}]{Lazio2001}
Lazio T. J.~W.,  Waltman E.~B.,  Ghigo F.~D.,  Fiedler R.~L.,  Foster R.~S.,   Johnston K.~J.,  2001, \mn@doi [The Astrophysical Journal Supplement Series] {10.1086/322531}, 136, 265

\bibitem[\protect\citeauthoryear{{Li}, {Li}, {Shu}  \& {Li}}{{Li} et~al.}{2024}]{Li2024}
{Li} H.,  {Li} R.-W.,  {Shu} P.,   {Li} Y.-Q.,  2024, \mn@doi [arXiv e-prints] {10.48550/arXiv.2404.01691}, \href {https://ui.adsabs.harvard.edu/abs/2024arXiv240401691L} {p. arXiv:2404.01691}

\bibitem[\protect\citeauthoryear{Liu}{Liu}{2021}]{Shu_Liu2021}
Liu S.,  2021, Regression: Book One, Series of Machine Learning with Scikit-Learn.
Independently published, \url {https://www.amazon.com/Regression-Book-Machine-Learning-Scikit-Learn/dp/B09BGHXBY6}

\bibitem[\protect\citeauthoryear{Lo, Farrell, Murphy  \& Gaensler}{Lo et~al.}{2014}]{Lo2014}
Lo K.~K.,  Farrell S.,  Murphy T.,   Gaensler B.~M.,  2014, \mn@doi [The Astrophysical Journal] {10.1088/0004-637X/786/1/20}, 786, 20

\bibitem[\protect\citeauthoryear{{Lorimer}, {Bailes}, {McLaughlin}, {Narkevic}  \& {Crawford}}{{Lorimer} et~al.}{2007}]{Lor2007}
{Lorimer} D.~R.,  {Bailes} M.,  {McLaughlin} M.~A.,  {Narkevic} D.~J.,   {Crawford} F.,  2007, \mn@doi [Science] {10.1126/science.1147532}, \href {https://ui.adsabs.harvard.edu/abs/2007Sci...318..777L} {318, 777}

\bibitem[\protect\citeauthoryear{Margutti et~al.,}{Margutti et~al.}{2018}]{Margutti2018}
Margutti R.,  et~al., 2018, \mn@doi [The Astrophysical Journal Letters] {10.3847/2041-8213/aab2ad}, 856, L18

\bibitem[\protect\citeauthoryear{{Merloni, A.} et~al.,}{{Merloni, A.} et~al.}{2024}]{Mer2024}
{Merloni, A.} et~al., 2024, \mn@doi [A&A] {10.1051/0004-6361/202347165}, 682, A34

\bibitem[\protect\citeauthoryear{{Mooley} et~al.,}{{Mooley} et~al.}{2018a}]{Moo2018a}
{Mooley} K.~P.,  et~al., 2018a, {A mildly relativistic wide-angle outflow in the neutron-star merger event GW170817}.
Nature

\bibitem[\protect\citeauthoryear{Mooley et~al.,}{Mooley et~al.}{2018b}]{Mooley2018c}
Mooley K.~P.,  et~al., 2018b, \mn@doi [The Astrophysical Journal Letters] {10.3847/2041-8213/aaeda7}, 868, L11

\bibitem[\protect\citeauthoryear{{Murphy} et~al.,}{{Murphy} et~al.}{2021}]{Mur2021}
{Murphy} T.,  et~al., 2021, arXiv e-prints, \href {https://ui.adsabs.harvard.edu/abs/2021arXiv210806039M} {p. arXiv:2108.06039}

\bibitem[\protect\citeauthoryear{Newman, King, Elvira, de Valpine, McCrea  \& Morgan}{Newman et~al.}{2023}]{New2023}
Newman K.,  King R.,  Elvira V.,  de Valpine P.,  McCrea R.,   Morgan B. J.~T.,  2023, \mn@doi [Methods in Ecology and Evolution] {https://doi.org/10.1111/2041-210X.13833}, 14, 26

\bibitem[\protect\citeauthoryear{Osten, Hawley, Allred, Johns-Krull  \& Roark}{Osten et~al.}{2005}]{Ost2005}
Osten R.~A.,  Hawley S.~L.,  Allred J.~C.,  Johns-Krull C.~M.,   Roark C.,  2005, The Astrophysical Journal, 621, 398

\bibitem[\protect\citeauthoryear{Paninski}{Paninski}{2010}]{Pan2010}
Paninski 2010, A new look at state-space models for neural data.
PMC

\bibitem[\protect\citeauthoryear{Pesaran}{Pesaran}{2015}]{pes2015}
Pesaran M.~H.,  2015, in , {Time Series and Panel Data Econometrics}.
Oxford University Press (\mn@eprint {} {https://academic.oup.com/book/0/chapter/364200282/chapter-pdf/44352583/acprof-9780198736912-chapter-4.pdf}), \mn@doi{10.1093/acprof:oso/9780198736912.003.0004}, \url {https://doi.org/10.1093/acprof:oso/9780198736912.003.0004}

\bibitem[\protect\citeauthoryear{{Powell}, {Sun}, {Gereb}, {Lasky}  \& {Dollmann}}{{Powell} et~al.}{2023}]{Pow2023}
{Powell} J.,  {Sun} L.,  {Gereb} K.,  {Lasky} P.~D.,   {Dollmann} M.,  2023, \mn@doi [Classical and Quantum Gravity] {10.1088/1361-6382/acb038}, \href {https://ui.adsabs.harvard.edu/abs/2023CQGra..40c5006P} {40, 035006}

\bibitem[\protect\citeauthoryear{{Predehl, P.} et~al.,}{{Predehl, P.} et~al.}{2021}]{Pre2021}
{Predehl, P.} et~al., 2021, \mn@doi [A&A] {10.1051/0004-6361/202039313}, 647, A1

\bibitem[\protect\citeauthoryear{{Rau}}{{Rau}}{2019}]{Rau2019}
{Rau} A.,  2019, Southern Horizons in Time-Domain Astronomy.
{Griffin}, R. Elizabeth

\bibitem[\protect\citeauthoryear{{Rau}}{{Rau}}{2023}]{Bla2023}
{Rau} A.,  2023, JWST Observations of the Extraordinary GRB 221009A Reveal an Ordinary Supernova Without Signs of $r$-Process Enrichment in a Low-Metallicity Galaxy (\mn@eprint {arXiv} {2308.14197})

\bibitem[\protect\citeauthoryear{{Rowlinson} et~al.,}{{Rowlinson} et~al.}{2019}]{Row2019}
{Rowlinson} A.,  et~al., 2019, \mn@doi [Astronomy and Computing] {10.1016/j.ascom.2019.03.003}, \href {https://ui.adsabs.harvard.edu/abs/2019A&C....27..111R} {27, 111}

\bibitem[\protect\citeauthoryear{Russell et~al.,}{Russell et~al.}{2024}]{Russell_2024}
Russell T.~D.,  et~al., 2024, \mn@doi [Nature] {10.1038/s41586-024-07133-5}, 627, 763–766

\bibitem[\protect\citeauthoryear{Sarkar, Gupta, Chitnis  \& Wiita}{Sarkar et~al.}{2020}]{Sark2020}
Sarkar A.,  Gupta A.~C.,  Chitnis V.~R.,   Wiita P.~J.,  2020, \mn@doi [Monthly Notices of the Royal Astronomical Society] {10.1093/mnras/staa3211}, 501, 50

\bibitem[\protect\citeauthoryear{Sharma, Swayne  \& Obimbo}{Sharma et~al.}{2016}]{Sharma2016}
Sharma S.,  Swayne D.,   Obimbo C.,  2016, \mn@doi [Energy, Ecology and Environment] {10.1007/s40974-016-0011-1}, 1

\bibitem[\protect\citeauthoryear{Shumway \& Stoffer}{Shumway \& Stoffer}{2017}]{shum2017}
Shumway R.~H.,  Stoffer D.~S.,  2017, Time Series Analysis and Applications.
Springer, \url {https://link.springer.com/book/10.1007/978-3-319-52452-8}

\bibitem[\protect\citeauthoryear{{Sunyaev, R.} et~al.,}{{Sunyaev, R.} et~al.}{2021}]{Sun2021}
{Sunyaev, R.} et~al., 2021, \mn@doi [A&A] {10.1051/0004-6361/202141179}, 656, A132

\bibitem[\protect\citeauthoryear{{Swinbank} et~al.,}{{Swinbank} et~al.}{2015}]{Swi2015}
{Swinbank} J.~D.,  et~al., 2015, \mn@doi [Astronomy and Computing] {10.1016/j.ascom.2015.03.002}, \href {https://ui.adsabs.harvard.edu/abs/2015A&C....11...25S} {11, 25}

\bibitem[\protect\citeauthoryear{Sánchez et~al.,}{Sánchez et~al.}{2019}]{San2019}
Sánchez B.,  et~al., 2019, \mn@doi [Astronomy and Computing] {https://doi.org/10.1016/j.ascom.2019.05.002}, 28, 100284

\bibitem[\protect\citeauthoryear{Templeton \& Karovska}{Templeton \& Karovska}{2009}]{Tem2009}
Templeton M.~R.,  Karovska M.,  2009, \mn@doi [The Astrophysical Journal] {10.1088/0004-637x/691/2/1470}, 691, 1470

\bibitem[\protect\citeauthoryear{Triantafyllopoulos}{Triantafyllopoulos}{2021}]{Tri2021}
Triantafyllopoulos K.,  2021, The State Space Model in Finance.
Springer International Publishing, Cham, pp 341--402, \mn@doi{10.1007/978-3-030-76124-0_7}, \url {https://doi.org/10.1007/978-3-030-76124-0_7}

\bibitem[\protect\citeauthoryear{Troja et~al.,}{Troja et~al.}{2019}]{Troja_2019}
Troja E.,  et~al., 2019, \mn@doi [Monthly Notices of the Royal Astronomical Society] {10.1093/mnras/stz2248}

\bibitem[\protect\citeauthoryear{Truong, Oudre  \& Vayatis}{Truong et~al.}{2020}]{Tru2020}
Truong C.,  Oudre L.,   Vayatis N.,  2020, \mn@doi [Signal Processing] {10.1016/j.sigpro.2019.107299}, 167, 107299

\bibitem[\protect\citeauthoryear{Tusell}{Tusell}{2008}]{Tus2008}
Tusell F.,  2008, \mn@doi [Journal of the Royal Statistical Society Series A: Statistics in Society] {10.1111/j.1467-985X.2008.00538_3.x}, 171, 756

\bibitem[\protect\citeauthoryear{Wang et~al.,}{Wang et~al.}{2021}]{Wang2021}
Wang Z.,  et~al., 2021, \mn@doi [The Astrophysical Journal] {10.3847/1538-4357/ac2360}, 920, 45

\bibitem[\protect\citeauthoryear{Wei}{Wei}{2019}]{William2019}
Wei W. W.~S.,  2019, Multivariate Time Series Analysis and Applications.
Wiley, \url {https://www.wiley.com/en-us/Multivariate+Time+Series+Analysis+and+Applications-p-9781119502852}

\bibitem[\protect\citeauthoryear{Zackay, Ofek  \& Gal-Yam}{Zackay et~al.}{2016}]{Zac2016}
Zackay B.,  Ofek E.~O.,   Gal-Yam A.,  2016, \mn@doi [The Astrophysical Journal] {10.3847/0004-637X/830/1/27}, 830, 27

\makeatother
\end{thebibliography}






\end{document}